\documentclass[useAMS,usenatbib]{mn2e}
\pdfoutput=1
\usepackage{times}

\usepackage{graphicx}
\usepackage{epstopdf} 
\usepackage{amssymb}
\usepackage{natbib}
\usepackage{hyperref}
\usepackage[latin1]{inputenc}
\newcommand{\kms} {$\mbox{km s}^{-1}$}
\newcommand{\kmskpc} {$\mbox{km s}^{-1}\;\mbox{kpc}^{-1}$}
\newcommand{\Msun} {$\mbox{M}_{\sun}$}

\def\spose#1{\hbox to 0pt{#1\hss}}
\def\lta{\mathrel{\spose{\lower 3pt\hbox{$\sim$}}
    \raise 2.0pt\hbox{$<$}}}
\def\gta{\mathrel{\spose{\lower 3pt\hbox{$\sim$}}
    \raise 2.0pt\hbox{$>$}}}
\newdimen\hssize
\newdimen\hdsize
\title[Nuclear fueling in a sub-parsec resolution galaxy simulation]
{The interplay between a galactic bar and a supermassive black hole: nuclear fueling in a sub-parsec resolution galaxy simulation}
\author
[Eric Emsellem et al.]{\parbox{\textwidth}{Eric Emsellem,$^{1,2}$\thanks{E-mail: eric.emsellem@eso.org \texttt{}}
Florent Renaud$^{3}$, Frédéric Bournaud$^{3}$, Bruce Elmegreen$^{4}$, Françoise Combes$^{5}$,
Jared M. Gabor$^{3}$}\vspace{0.4cm}\\
\parbox{\textwidth}{$^{1}$European Southern Observatory, Karl-Schwarzschild-Str. 2, 85748 Garching, Germany\\
$^{2}$Universit\'e Lyon 1, Observatoire de Lyon, Centre de Recherche Astrophysique de Lyon \\ \hspace*{0.5cm} and Ecole Normale Sup\'erieure de Lyon, 9 avenue Charles Andr\'e, F-69230 Saint-Genis Laval, France\\
$^{3}$Laboratoire AIM Paris-Saclay, CEA/IRFU/SAp -- CNRS -- Universit\'e Paris Diderot, 91191 Gif-sur-Yvette Cedex, France \\
$^{4}$IBM Research Division, T.J. Watson Research Center, 1101 Kitchawan Road, Yorktown Heights, NY 10598, USA \\
$^{5}$Observatoire de Paris, LERMA, CNRS: UMR8112, 61 Av. de l'Observatoire, 75014 Paris, France 
}}
\begin{document}
\maketitle
%
%
\begin{abstract}
    We study the connection between the large-scale dynamics and the gas fueling toward a central black hole via the analysis of a Milky Way-like simulation at sub-parsec resolution. This allows us to follow a set of processes at various scales (e.g., the triggering of inward gas motion towards inner resonances via the large-scale bar, the connection to the central black hole via mini spirals) in a self-consistent manner. This simulation provides further insights on the role of shear for the inhibition of star formation within the bar in regions with significant amount of gas. We also witness the decoupling of the central gas and nuclear cluster from the large-scale disc, via interactions with the black hole. This break of symmetry in the mass distribution triggers the formation of gas clumps organised in a time-varying 250~pc ring-like structure, the black hole being offset by about 70~pc from its centre. Some clumps form stars, while most get disrupted or merge. Supernovae feedback further creates bubbles and filaments, some of the gas being expelled to 100~pc or higher above the galaxy plane. This helps remove angular momentum from the gas, which gets closer to the central dark mass. Part of the gas raining down is being accreted, forming a 10~pc polar disc-like structure around the black hole, leading to an episode of star formation. This gives rise to multiple stellar populations with significantly different angular momentum vectors, and may lead to a natural intermittence in the fueling of the black hole.    
\end{abstract}

\section{Introduction\label{sec:intro}}

One of the most common but daunting challenges for simulating astrophysical objects is to be able to 
properly treat various physical processes and their associated scales simultaneously. 
The onset and impact of gas fueling within a few hundreds of parsecs driven by a kpc-scale bar 
in a galaxy \citep{Roberts1979, Shlosman1989, Athanassoula1992, Jogee2005, 
Haan2009, Cisternas2013, Combes2014, Garcia-Burillo2014} is an excellent illustration of 
a problem which requires pushing our models and facilities to their limits. 
Assuming that we ignore the interactions with the galactic environment and the associated 
infall and outflows of material, angular momentum, and energy, probing the formation and evolution
of the bar requires to follow the large-scale dynamics, hence the disc scale with tens of kiloparsecs,
and its potential coupling with star formation and feedback, therefore
down to sub-parsec regions if we are to resolve the associated local quasi-isotropic
turbulent motions \citep{Bournaud2010}.

In a gas-rich galaxy, we expect a rapidly time-varying system with strong coupling from processes such
as global and local instabilities, turbulence (3D and inverse 2D) cascades and star formation 
\citep{Elmegreen1993,Padoan2002,Levine2008,Bournaud2010,Padoan2011,Hopkins2010}. We also expect the interplay between
the underlying frequencies set up by the tumbling and evolving potential (pattern speed of the bar and spirals) 
and the local frequencies (e.g., epicycle) to have a strong impact on how and where the gas exchanges angular
momentum or forms stars \citep[see e.g.][]{Combes2001,Jogee2006}. 
Resonances will thus shape the orbital structure of the galaxy and are therefore critical ingredients
to understand the overall evolution \citep{Binney1991,Athanassoula1992,Wada1995,Englmaier2000,Stark2004}. 
In that context, the presence (or absence) of a supermassive black hole
at the centre of the galaxy will play a prominent role, as it significantly modifies both the profile and
shape of the potential in its close environment \citep{Englmaier2000, Maciejewski2004}, 
and consequently the presence and location of the inner
resonances (e.g., the Inner Lindblad Resonance), and also act as a massive particle among a sea of stars and
gas clouds \citep{Merritt2010}. The feedback from the active nucleus, when included, could also affect the evolution of the system
both locally and globally \citep{Haas2013, Gabor2014, Choi2014}. Molecular gas structures such as spirals and outflows
in the nuclear regions of three local barred galaxies have been mapped recently at high angular resolution; 
they suggest the presence of high torques that can drive accretion to a central black hole 
\citep{Combes2013, Combes2014, Garcia-Burillo2014}.

In the present paper, we are making use of a Milky-Way like simulation to probe the coupling
between the gas fueling driven by a large-scale bar and the central region surrounding a supermassive black
hole. For such a study, we focus on an isolated galactic system to further understand the physical ingredients
which may indeed associate the large-scale evolution in the disc with the very central structures. We 
also focus on the dynamical processes, as well as on the associated star formation and its 
feedback, thus ignoring the more global cosmological context or the active galactic nucleus' (AGN) feedback.

In the next Section, we briefly describe the ingredients for such a simulation. In Section 3, we
present the first results, following on the formation and evolution of the bar and the triggered gas fueling
leading to the emergence of complex central structures in the close environment of the black hole. In Section
4, we briefly discuss these results, adding some perspectives for follow-up studies. A summary of the
structures we observe in the simulation is provided in Fig.~\ref{fig:Sketch}.

\section{The Milky-Way simulation}
\label{sec:methods}

We use the Milky Way-like galaxy simulation presented in
\cite[][thereafter R+13]{Renaud2013}, and of which the main aspects are 
summarized below. A disc galaxy is modeled in three
dimensions, with live dark matter halo and stellar components rendered with 60 millions particles. In
addition, a super massive black hole of $4 \times 10^6$~\Msun\ is initially set at the center of the galaxy:
the black hole is treated as a single massive particle with no direct accretion, growth or 
associated feedback. For theses
components (dark matter, primordial stars and super-massive black hole), the gravitational potential is
computed using a particle-mesh technique with the resolution (i.e. softening) of 3~pc, as to limit the 
noise due to individual particles. The gaseous component
is setup on a grid, which follows the Adaptive Mesh Refinement (AMR) technique with a refinement strategy
based on local volume density and Jeans length. The smallest cell size spans 0.05~pc in the most refined
regions, which is thus the minimum softening scale for both the gas and the stars formed during the
simulation.

The equations of motions and hydrodynamics are solved by the RAMSES code 
\citep{Teyssier2002a}. The gas follows a piecewise polytropic equation of state (EoS) 
fitting the heating/cooling equilibrium \citep[see][and references therein]{Kraljic2014}. 
A Jeans polytrope sets a pressure floor in the most refined volumes, to prevent artificial fragmentation. 
We refer the reader specifically to Sections 2.2, 2.3 and 4.2 of R+13 and \cite{Kraljic2014} 
for further discussions on this and specific EoS 
\citep[see][for a few other implementation schemes]{Robertson2008,Tasker2008,Dobbs2011,Bonnell2013}.
The resulting gas density probability distribution function (PDF) \citep[][and
references therein]{McKee2007} in the present simulation follows a classic log-normal shape \citep{Nordlund1999,Padoan2002} with an additional
few percents of the mass in a power-law tail at high density (R+13, \cite{Choi2013}),
as expected from gravity and observed in real molecular clouds and galaxies
\citep{Lombardi2010,Druard2014}. While changes in the AMR grid refinement can locally bias the velocity dispersion, 
the density and velocity power spectra are thus clearly dominated by a single turbulence cascade with a 
well-identified injection scale at the average Jeans length \citep[][R+13]{Bournaud2010}.
The very high resolution of the present simulation
allows to resolve the turbulent cascade with a relastic power spectrum (R+13, \cite{Combes2013}) 
and density distribution \citep{Druard2014} down to the parsec scale. 

The simulation comprises
conversion of gas into stellar particles (down to 160~\Msun) where the volume density $\rho_0$ exceed 2000~cm$^{-3}$,
assuming that the local star formation rate depends on the free-fall time and 
with the star formation efficiency set at 3\% (R+13). This recipe does not take into account additional
physics that may impact on the formation of molecules, and we thus rely on the equation of state to follow
the cloud collapse, the low temperatures and the high density which triggers the formation of new stars.
These newly formed ``stars'' are evolved on the AMR grid, i.e. with gravitational softening down to 0.05~pc,
much smaller than the dark matter and primordial stellar components. The implementation of stellar feedback
includes photoionisation through heating, radiative pressure via injection of momentum, and supernova
explosions in the kinetic form (see R+13 for more details). A more thorough study of the impact of resolution
or metallicity in such simulations was conducted by \cite{Kraljic2014}, who have shown
that the artificial density threshold $\rho_0$ does not tune the efficiency of star formation,
which mostly depends on the turbulence level \citep[e.g., Mach number, see][]{Klessen2000,Li2004,Audit2010,Renaud2012}.

The artificial viscosity induced by the refinement process is also not expected to induce e.g., spurious
fuelling even close to the potential resonances in such a simulation. Introducing a new refinement level induces some
additional dispersion within the new cells which are created. That dispersion is itself not a true turbulence term. 
The power spectrum is in fact dominated by the injection of turbulence from the large scale, and the higher resolution in these new cells also 
allow the gas to cool further. The thermal energy is thus replaced by kinetic energy but this does not induce turbulence per se. 
A simple calculation pertaining to the present simulation can provide some further estimate of this process. 
Assuming a mean free path of one cell, with 
velocities associated with sound waves, we can estimate an upper limit of the (viscous) 
acceleration in an homogeneous medium as $a_{visc} = (c_s \, dv/dr) / 2$
(with $c_s$ the sound speed). At the resonance ($\sim 500$~pc), the gas is dense, 
with densities around 100~H/cc, hence $dv/dr$ is of the order of 50~km.s$^{-1}$.kpc$^{-1}$
which leads to $a_{visc} \sim 0.03$~pc.Myr$^{-2}$ (for $c_s \sim 1$~\kms). This estimate would significantly
decrease if we were to consider the realistic filling factor of real gas clouds (and not an homogeneous medium). 
The energy ($E$) dissipation rate for the turbulence per unit mass 
at the resonance is then $dE/dt = 3/2 \, \sigma^2 / \tau$ where $\sigma \sim 10$~\kms\ is the one-dimensional turbulence (root mean square) 
and $\tau$ the free-fall time ($\sim 5$~Myr for densities around 100~H/cc). We then get
the acceleration associated with the turbulence via $a_{turb} = 3\, \sigma^2 /  \left(2 \tau \cdot V\right)$. With $V \sim \,100$~\kms\ at the resonance, 
this leads to $a_{turb} \sim 0.3$~pc.Myr$^{-2}$. Hence, artificial viscosity would have a negligible role in the triggering of gas fuelling,
even close to the resonances.

The results presented in this paper focus on the fueling due to the stellar bar which forms during the
simulation, and the associated evolution of the inner region, down to the vicinity of the black hole.
We have conducted two runs at two different resolutions following the same initial conditions.  
The first simulation (or run) is as described in R+13 and reaches its maximum resolution of 0.05~pc
(refinement level 21) in a number of very dense regions. Within the bar, that maximum resolution is only
reached late in the simulation just before its end time at about $t=800$~Myr. This motivated the
launch of a second run at slightly coarser resolution (cells of about 1~pc, 
with a maximum level of refinement of 17) to pursue the evolution of the central region for an additional
30~Myr (i.e., about the dynamical time-scale at a radius of 1~kpc or 10 dynamical time-scales at 100~pc). 

In the following, all measurements and maps derived
from snapshots prior to $t=800$~Myr are therefore using the original simulation (R+13),
with the others extracted from the simulation at 1~pc resolution (for the gaseous component).
Note that star formation was turned on at $t\sim745$~Myr in the simulation as to 
avoid gas being prematurely consumed.
Details on the implemented recipes for star formation, stellar feedback (photoionisation, 
radiative pressure, supernova explosions) are described in R+13. Adding on AGN feedback 
would significantly impact on the distribution, kinematics and physical status of the gas, 
specifically for the close environment of the black hole \citep{Ciotti1997,Haehnelt1998,Silk1998,Kauffmann2000}. For the present simulations, however,
we do not include the potential feedback from an active galactic nucleus (AGN), thus focusing on a time window
($t \sim 750 - 830$~Myr) when we consider that the AGN itself is quiet 
(or in an ``off-state''). This is partly justified by the assumption that AGN have low duty cycle at
low redshift and for black holes of a few $10^6$~\Msun\ \citep{Haehnelt1993, Wang2009,Shankar2010, Shankar2013}
and by the short time range we are considering. More importantly, it allows us to narrow down
our study to probe the interplay between the dynamical evolution and the effect of star formation 
\citep[similarly to e.g.,][]{Levine2008,Hopkins2010}.
Turning on the AGN in such a simulation would be paramount to understand any potential fueling cycle
starting from the large-scale down to the vicinity of the black hole, 
and such an implementation has already been included in RAMSES by a direct calculation of the Bondi accretion
rate \citep[see][and references therein]{Teyssier2011,Gabor2014}.
It would nevertheless require to probe various feedback schemes, and triggering mechanisms, 
which is beyond the scope of the present paper.


\section{Results}
\label{sec:results}

In the following, we roughly reconstruct the chronology of events as witnessed in the simulation.
The large-scale bar forms, fuels gas towards the centre and the inner Lindblad resonance, 
creating a spiral-like structure and a ring, with part of this gas spiraling down to the inner few parsecs 
around the black hole, triggering star (cluster) formation. The new stellar cluster and the 
black hole are bound to each other, 
and wander away from the center of gravity of the older stellar population. 
This symmetry breaking event speeds up the fragmentation of the ring, 
triggering further localised starbursts in clumps as well as
in the close vicinity of the black hole. Part of the gas is expelled tens of parsecs away 
from the galactic plane via stellar feedback, 
and is accreted back within an inclined disc of about 10~pc around the black hole.
That out-of-the-plane accretion induces further star formation in the close vicinity of the central dark mass.
These events are detailed in the following Sections.

\subsection{Bar formation and gas fueling}
\begin{figure}
\centering
\includegraphics[width=\columnwidth]{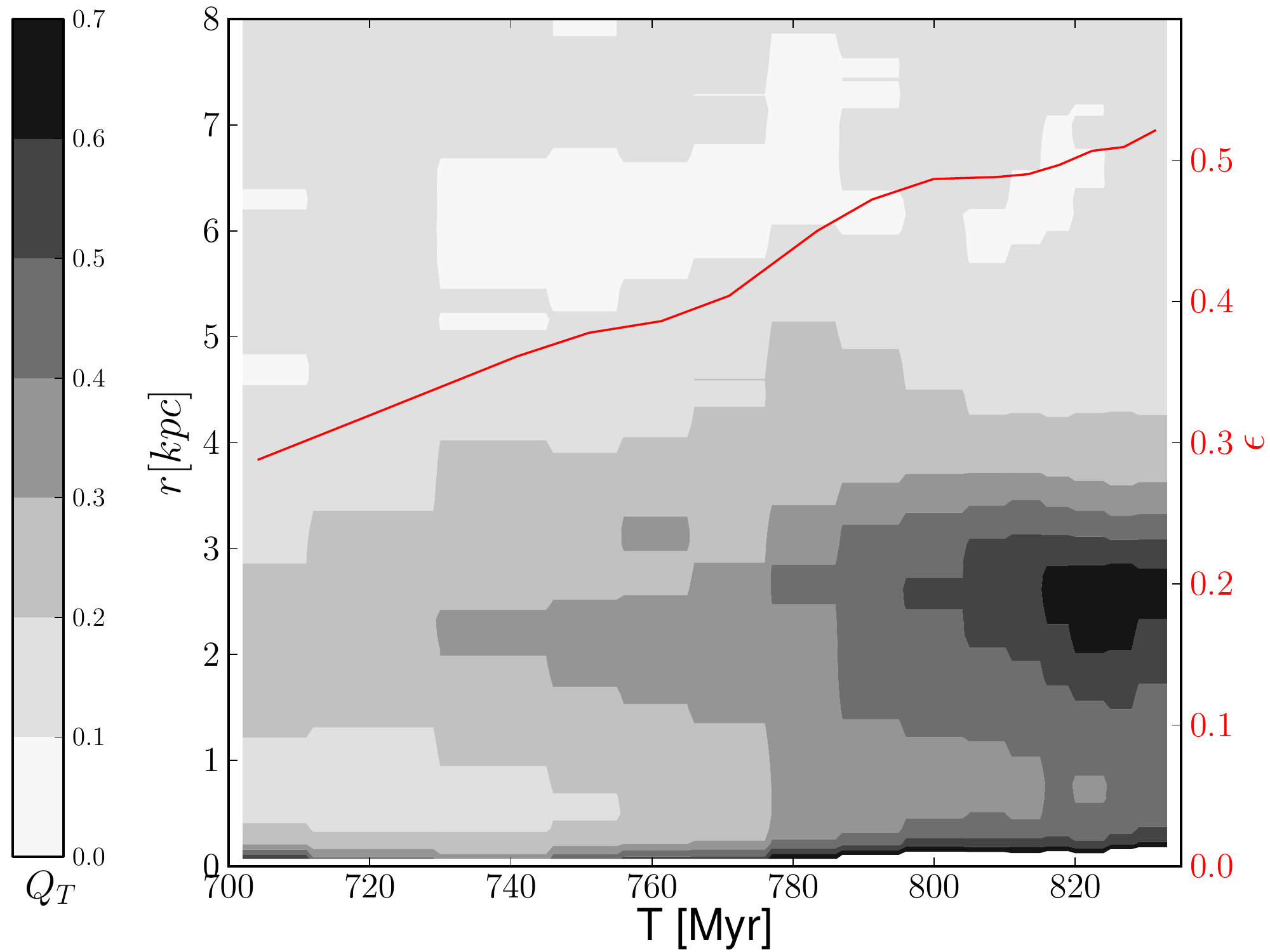}
\caption{Tangential to radial force ratio $Q_T(r)$ (left $y$ axis) for different snapshots of the 
simulation presented with respect to time (in Myr), and radius
(left label in kpc). The amplitude of $Q_T$ is colour-coded 
and the correspondence is provided in the colourbar at the left of the plot. The time variation of the
ellipticity of the bar is provided as a red solid line (labels on the right).}
\label{fig:QT}
\end{figure}

\subsubsection{The bar formation}
\begin{figure}
\centering
\includegraphics[width=\hsize]{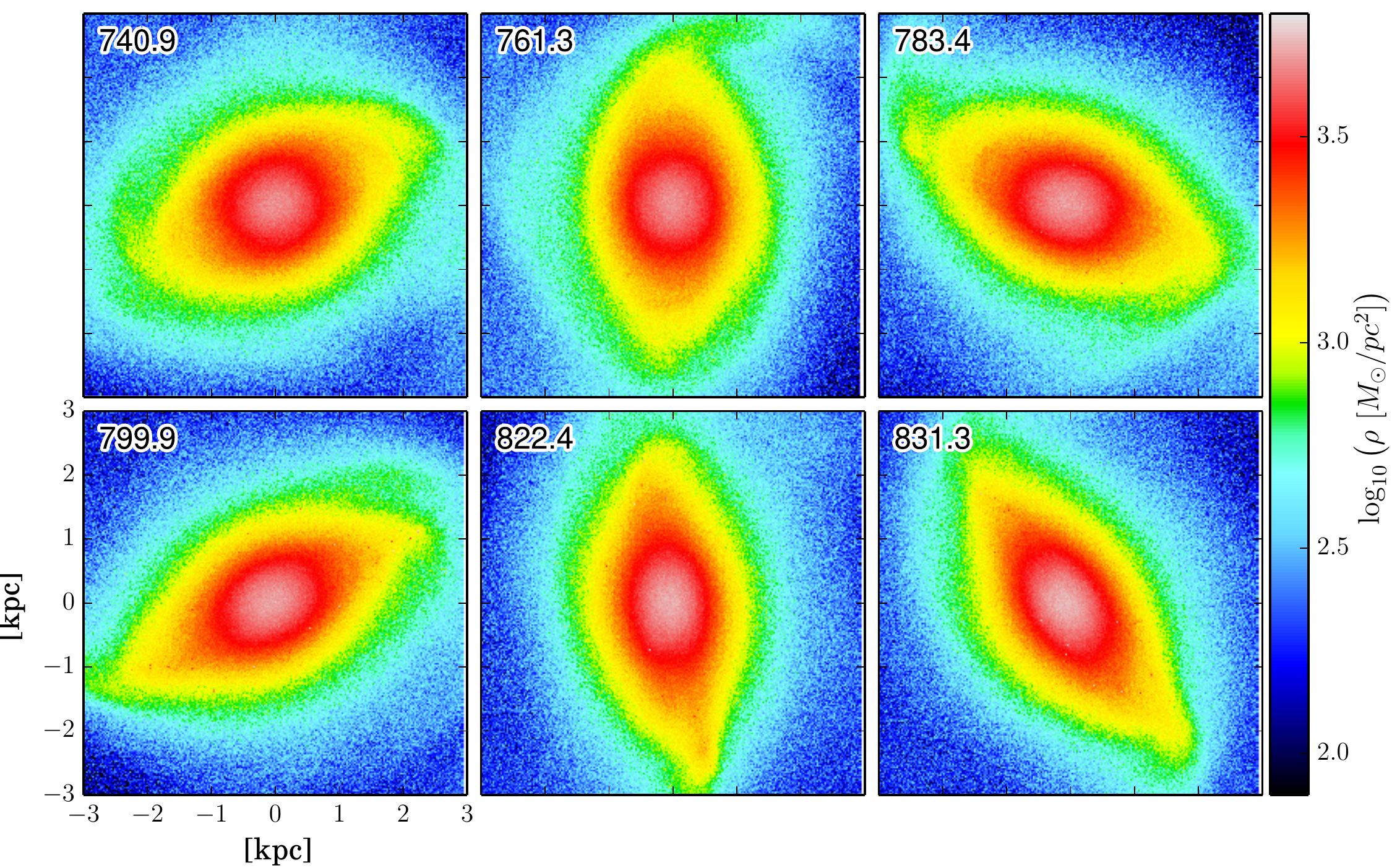}
\caption{Face-on view of the surface (mass) stellar density within 3~kpc of the simulation for 6 snapshots showing the evolution of
the large-scale stellar bar. The snapshot time (in Myr) is indicated in the upper left corner of each panel.}
\label{fig:6starpanels}
\end{figure}
\begin{figure*}
\centering
\includegraphics[width=\hsize]{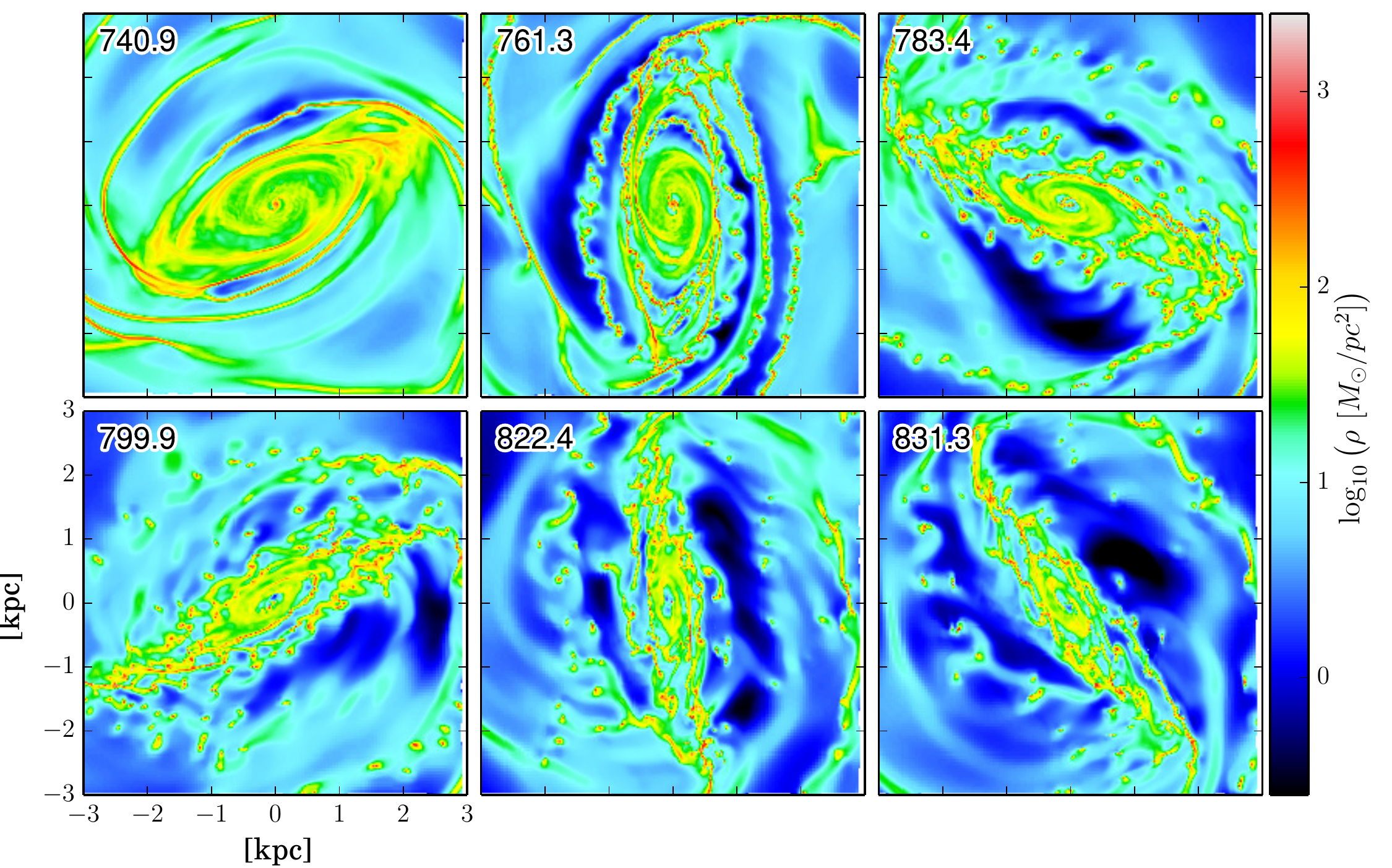}
\includegraphics[width=\hsize]{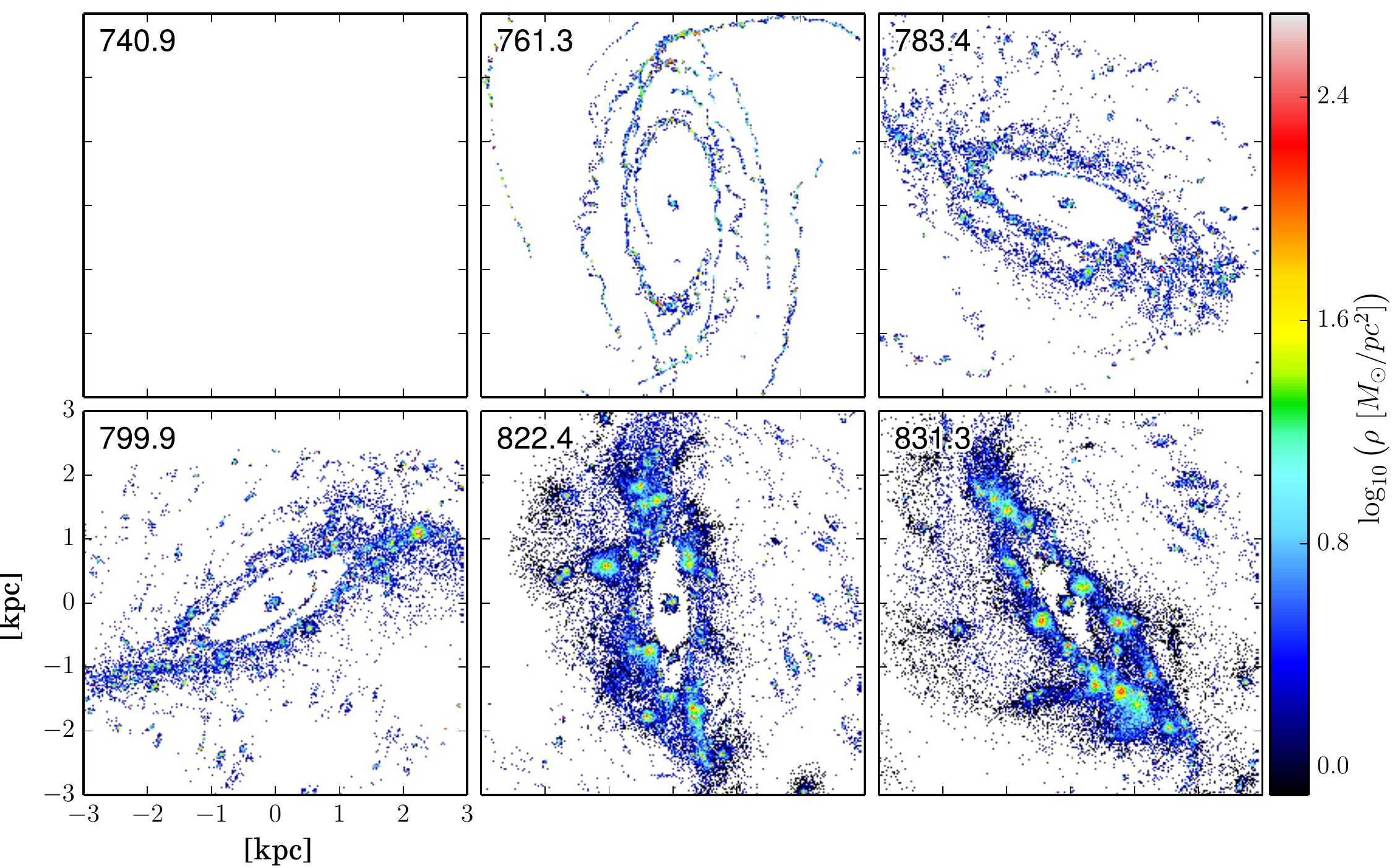}
\caption{Same as in Fig.~\ref{fig:6starpanels} but for the gas (upper 6 panels) and 
all stars formed during the simulation (lower 6 panels). 
The top left panel ($t=740.9$~Myr) stays blank as the star formation was not yet turned on at that
stage of the simulation (see Sect.~\ref{sec:methods}). }
\label{fig:6gaspanels}
\end{figure*}

The bar is already formed at $t=700$~Myr, and dominates the old stellar component within the central 3~kpc 
(R+13). It stills grows significantly until the end
of the simulation run at 830~Myr: this is shown in Figure~\ref{fig:QT} by the time variation of the force ratio 
$Q_T(r) = \left| F_T(r,\theta) \right|_{max} / \left< \left| F_r(r,\theta)\right|\right>$ 
\citep{Combes1981, Buta2001}.
The maximum amplitude $Q_b = {\rm max}[Q_T(r)]$ goes from about 0.2 \citep[weak bar, see e.g.,][]{Eskridge2002, Buta2001, Buta2004, Buta2006} at 700~Myr 
to $\sim 0.65$ (strong bar) at 830~Myr, while the bar (two-dimensional) ellipticity increases from 0.3 to 0.5.
The maximum amplitude is reached at a radius of about 2.6~kpc.
As mentioned in R+13, the bar pattern speed $\Omega_b$ is about 58~\kmskpc, while the outer spirals are slower with
$\Omega_{\rm spiral} \sim 50$~\kmskpc. The bar corotation lies at $\sim 3.6$~kpc at 800~Myr.
The bar exerts strong torques which act on the gas by driving radial outflows and inflows.
As emphasised in R+13, inner resonances appear while the bars forms, with radii of $\sim 40$ and 450~pc.
In the rest of this paper, we focus on the evolution and central gas radial redistribution due to the bar, 
namely within the central $\sim 3$~kpc.

\subsubsection{The bar evolution}
\label{sec:overall}

With a pattern speed of $\sim 58$~\kmskpc, the bar rotates by about 45\degr\ every $\sim13$~Myr, as 
illustrated in Fig.~\ref{fig:6starpanels}: the ellipticity increase of the bar (Fig.~\ref{fig:QT}) thus occurs here in
less than a rotation period. The gas is significantly redistributed via the bar formation along spiral shocks
connecting at the end of the bar (Fig.~\ref{fig:6gaspanels}), which then follow several embedded banana-shaped thin structures: this seems to
follow the expected orbits around the Lagrangian points, as described by e.g., 
\cite{Athanassoula1992} and \cite{Maciejewski2002}. Three or four of these thin
overdensities meet near both ends of the stellar bar where the outer spirals start. They also curve towards
the centre resulting in a irregular ring-like structure at a radius of about 400~pc. The standard bi-symmetry
expected in idealised simulations \citep[see e.g.,][]{Maciejewski2002} is broken
here with superimposed $m=1$ and $m=3$ modes: these are visible
both in the stellar component (Fig.~\ref{fig:6starpanels}) as well as in the
three-arms gaseous overdensity at a radius of $\sim 500$~pc in Fig.~\ref{fig:6gaspanels}. 
At the end of the bar, gas densities go up to $\sim 10^4$~\Msun.pc$^{-2}$, 
while it is on average less than $10^3$~\Msun.pc$^{-2}$ in the inner spirals. More diffuse gas with densities of 
$[20 - 50]$~\Msun.pc$^{-2}$ is present within the
bar, filling the inter-arm regions. As the bar evolves and its ellipticity increases, the inner gaseous structure gets also
more elongated, but keeps a relatively regular appearance with only the spiral and shock lanes outside the bar
getting disrupted. Part of this disruption may be due to the density gradient and Rayleigh-Taylor instabilities 
\citep[see e.g.][]{Maciejewski2002} but also to the local shear and Kelvin-Helmholtz instabilities.

\subsubsection{The lack of star formation within the bar}
\label{sec:lack}

Within 3~kpc, stars form at an average rate of about 0.1~\Msun.yr$^{-1}$.
As expected, the star formation sites generally trace the overdensities described above. There is, however, a remarkable
exception: apart from a very centralised burst inside 40~pc (see Sect.~\ref{sec:centralkpc}), the region inside the bar 
(the inner kiloparsec or so) is devoid of new stars except in the central $\sim 100$~pc.
As illustrated in the bottom panel of Fig.~\ref{fig:6gaspanels},
star formation proceeds efficiently at the very edge of the bar, with the presence of 
many clumps following $x_1$ orbits along the bar. There are also two additional clump concentrations 
at the two ends of the bar. Within the bar itself, stars solely form either along a 
couple of thin curvy streams or at the very centre.

Considering the significant morphological changes of the bar (Fig.~\ref{fig:6starpanels}) and within the bar (Fig.~\ref{fig:6gaspanels}), 
it is difficult to directly assess the mass budget evolution in that region. A reasonable estimate can
still be obtained by following the variation of position angle 
and ellipticity of the density from the old stellar population,
and using that to measure the enclosed mass. The cumulated mass of gas and new stars within an elliptical
region with a major-axis of 500~pc (the average ellipticity of that region being 0.25) 
is almost constant with 1.8~$10^7$~\Msun, all new stars ($\sim
4\,10^6$~\Msun) being formed within the central 40~pc (see Sect.~\ref{sec:centralkpc}).
The fact that star formation is not triggered within the inner region of the bar despite gas densities being
above $10^3$~\Msun.pc$^{-2}$ is quite remarkable. This could originate either in processes preventing gas to collapse and form
stars within the bar region despite the high densities, or by the absence of 
processes needed to trigger star formation. 
Potential scenarios may involve cloud-cloud collisions as emphasised by \cite{Fujimoto2014},
which should be particularly important at the bar ends and in the outer spiral. To explain
the low star formation rate within the bar, intrinsic difference in the cloud properties (e.g.,
concentration) are invoked, but it does not directly point out the 
required physical origin for such differences \citep[but see][]{Kruijssen2014}.

\begin{figure}
\centering
\includegraphics[width=\columnwidth]{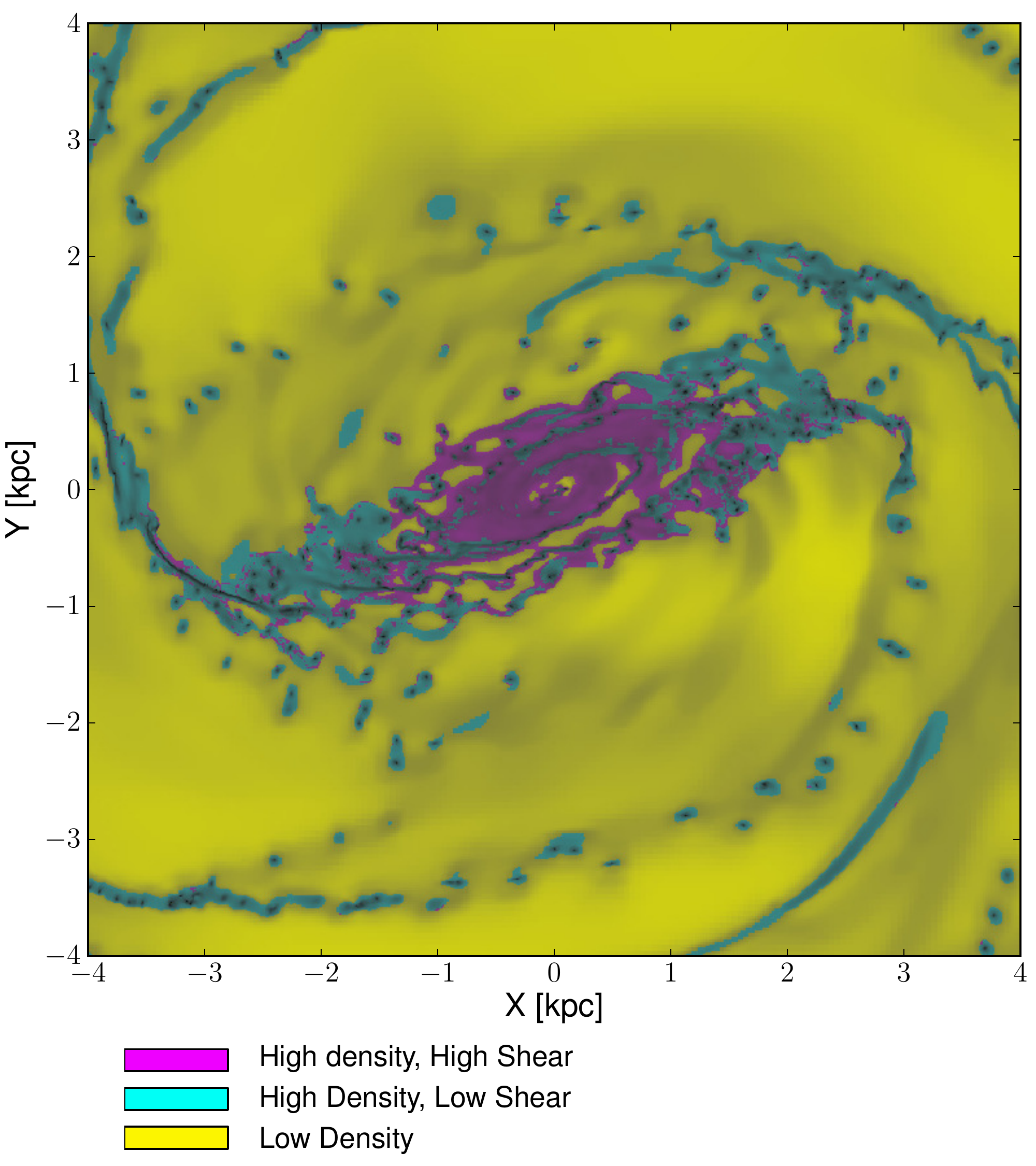}
\caption{Map of the 8 central kpc of the simulation showing both information about surface density and shear
at $t = 800$~Myr.
Low density gas is coded in yellow. High density gas with high (resp. low) shear is coded in magenta (resp.
cyan). The map clearly shows that high shear, high density gas is strictly confined to the inner region of the
bar. }
\label{fig:Shear}
\end{figure}

Stars thus form in the bar convergence region and principal shocks where there is a slight associated
increase of gas density (Fig.~\ref{fig:6gaspanels}). However, 
the difference between dense regions of gas within the bar which form stars or 
not is only marginally associated with the gas density itself: at a scale of 10~pc, the regions devoid
of star formation within the bar are of similar or even higher densities than the star forming ones at the end
of the bar or in the spiral. The present simulation distinctly shows 
such a well confined low star formation rate within a bar.
We should therefore have a closer look at the bar-driven gas flows to seek whether or not
the mentioned behaviour can be associated with the very specific (and complex) orbital and dynamical structure
induced by the bar. In a dynamical context, one of the local ingredients which could prevent gas clouds to
collapse is shear, namely forces which would tear down coherent gas structures, acting as a dynamically disruptive
agent on gas clouds. 

We probed this hypothesis by calculating the normalised shear at a scale of about 5~pc (refinement level 14). Namely, we
compute the quadratic sum of the non-diagonal terms of $F_{\alpha,\beta} = dV_{\alpha} / d\beta - dV_{\beta} /
d\alpha$ where $\alpha$ and $\beta$ represent two of the three cartesian coordinates ($x,y,z$), and then
calculate the mass-weighted average of that quantity normalised by the local gas self-gravity. In
Fig.~\ref{fig:Shear}, we present a map of the local normalised shear:
regions with low density are represented in yellow, and regions with high density are in cyan or magenta,
the former being associated with low shear while the latter emphasises high shear regions
(the gas density is also coded in the map brightness, higher density regions being darker). 
High density gas, namely with a volume density larger than about 2000~cm${-3}$, corresponds to 
the power-law tail of the probability density function (see R13), hence should thus be self-gravitating
and is expected to collapse and form stars if no other mechanism opposes it.

Small star forming clumps around or outside the bar are all in regions of low shear.
There is in fact an excellent correspondence between the location of high density clumps at low shear and 
the regions where the new stars form (see Fig.~\ref{fig:6gaspanels}): the spiral arms are thus
all clearly flagged in cyan in Fig.~\ref{fig:Shear}, as well as the two ends of the bar.
The inner region of the bar, besides a small area around the very centre, 
is strikingly singled out in Fig.~\ref{fig:Shear} as having both high gas
densities and high shear. This is a strong hint that shear may be a significant actor in preventing star
formation to occur within the bar, except at the very centre which we probe further in the next Sections.

\subsection{The central kpc}

We now focus on the environment of the black hole within
the inner resonances where the gas is funnelled : the inner 100~pc is 
a region of active star formation (as opposed to its immediate surroundings within the bar).

\subsubsection{Evolution of the inner ring}
\label{sec:centralkpc}

\begin{figure*}
\centering
\includegraphics[width=\hsize]{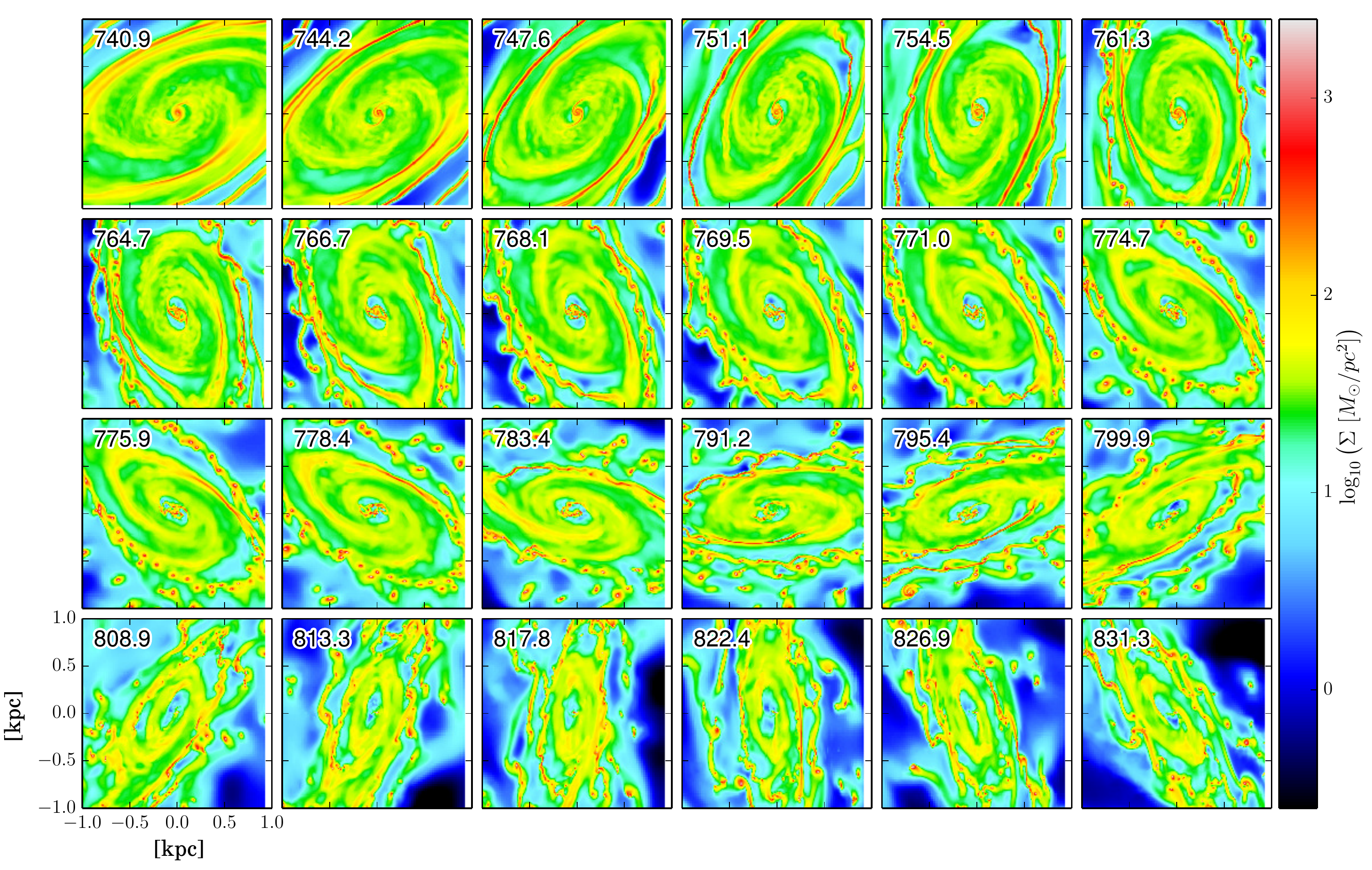}
\includegraphics[width=\hsize]{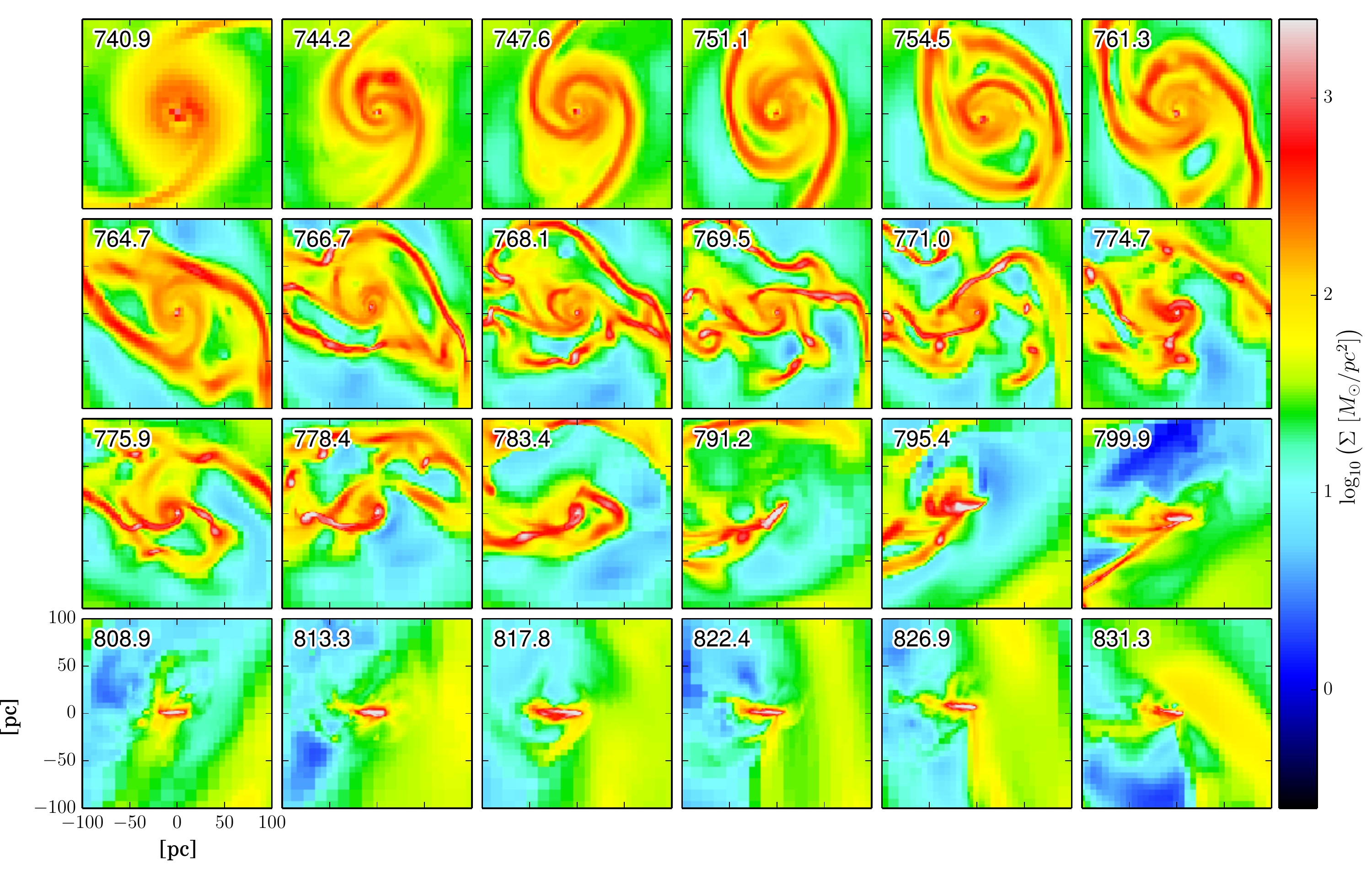}
\caption{Face-on view of the gas surface density distribution within the central 1~kpc (top panels) and 
100~pc (bottom panels).
Each set of 24 panels show about 90~Myr of evolution with an average of about 4~Myr between each panel.
The time (in Myr) of each snapshot is indicated at the top left corner of each panel. The black hole
is centred in all panels. The 20~pc size structure seen in the bottom panels after $\sim 800$~Myr is a
vertical disc structure (see Sect.~\ref{sec:clumps}).}
\label{fig:20Snaps}
\end{figure*}

The central structures associated with the bar inner Lindblad resonances are formed early with a clear
spiral-like morphology established by $t = 730$~Myr (Figs.~\ref{fig:6gaspanels} and \ref{fig:20Snaps}). 
The fueling rate is steady within 100~pc at a level of 0.02~\Msun.yr$^{-1}$ or less (Fig.~\ref{fig:EnclosedMass}).  
The two incoming arms connect to the central 40~pc disc-like
structure, respectively reaching surface densities of about 400 and 250~\Msun.pc$^{-2}$. 
As the gas accumulates, the arms then continue
down to a few parsecs distance from the black hole. Such structures may be reminiscent of the
gas distribution obtained in some hydrodynamical simulations in the presence of a bar
where the presence of an additional central mass can drive the propagation of a spiral down to the very centre, 
albeit with significantly lower amplitudes \citep[see e.g.][]{Englmaier2000}
or via an idealised scheme \citep{Maciejewski2004, Maciejewski2004a}.
These studies include a fixed (tumbling) potential while the present simulations includes live stellar and dark matter
components, star formation and feedback.

\begin{figure}
\centering
\includegraphics[width=\columnwidth]{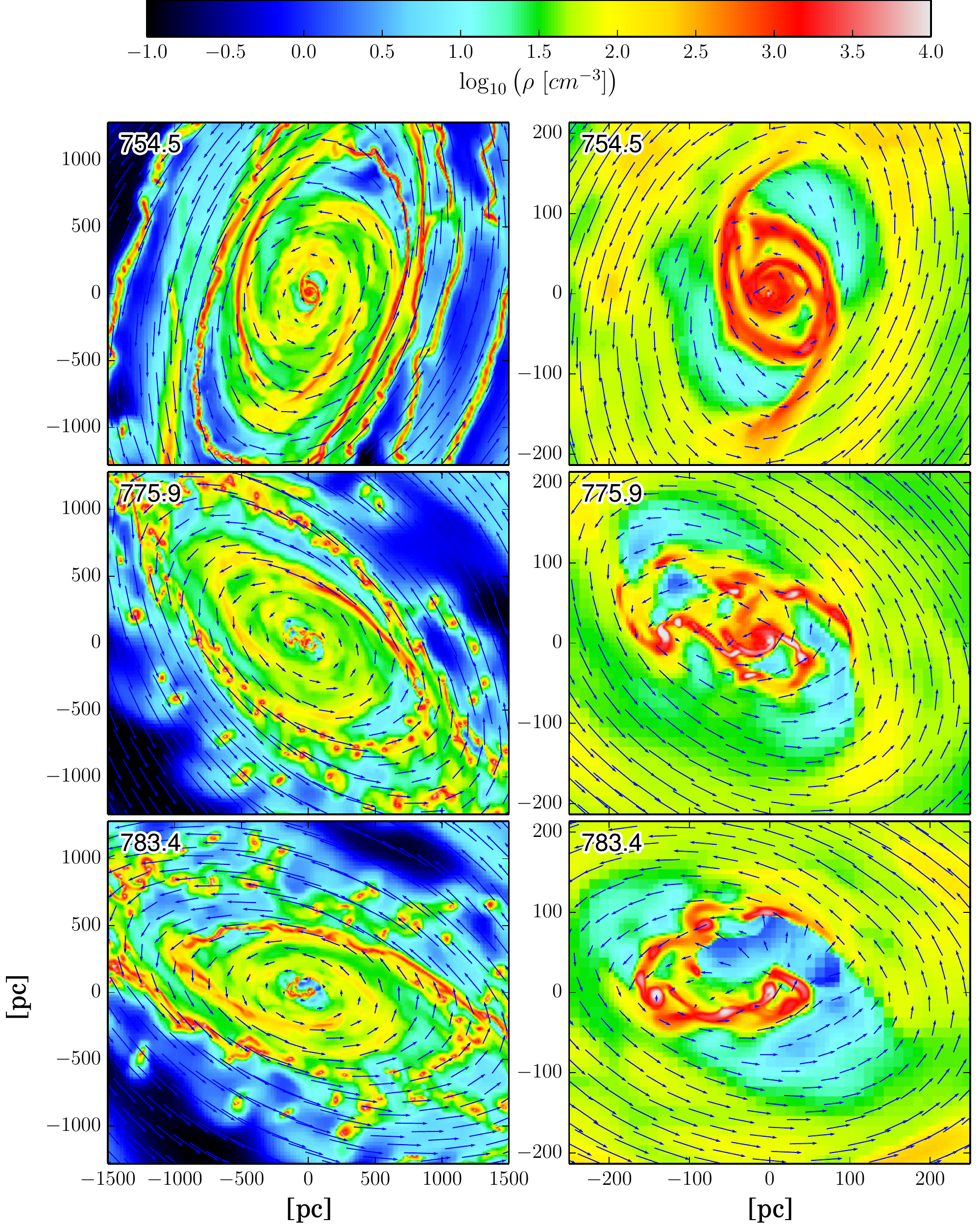}
\caption{Three snapshots of the gas volume density maps 
at 1~kpc (left panels) and 200~pc (right panels) scales, with the gas velocity
flow superimposed as blue arrows.}
\label{fig:3SnapsQuiver}
\end{figure}

\begin{figure}
\centering
\includegraphics[width=\columnwidth]{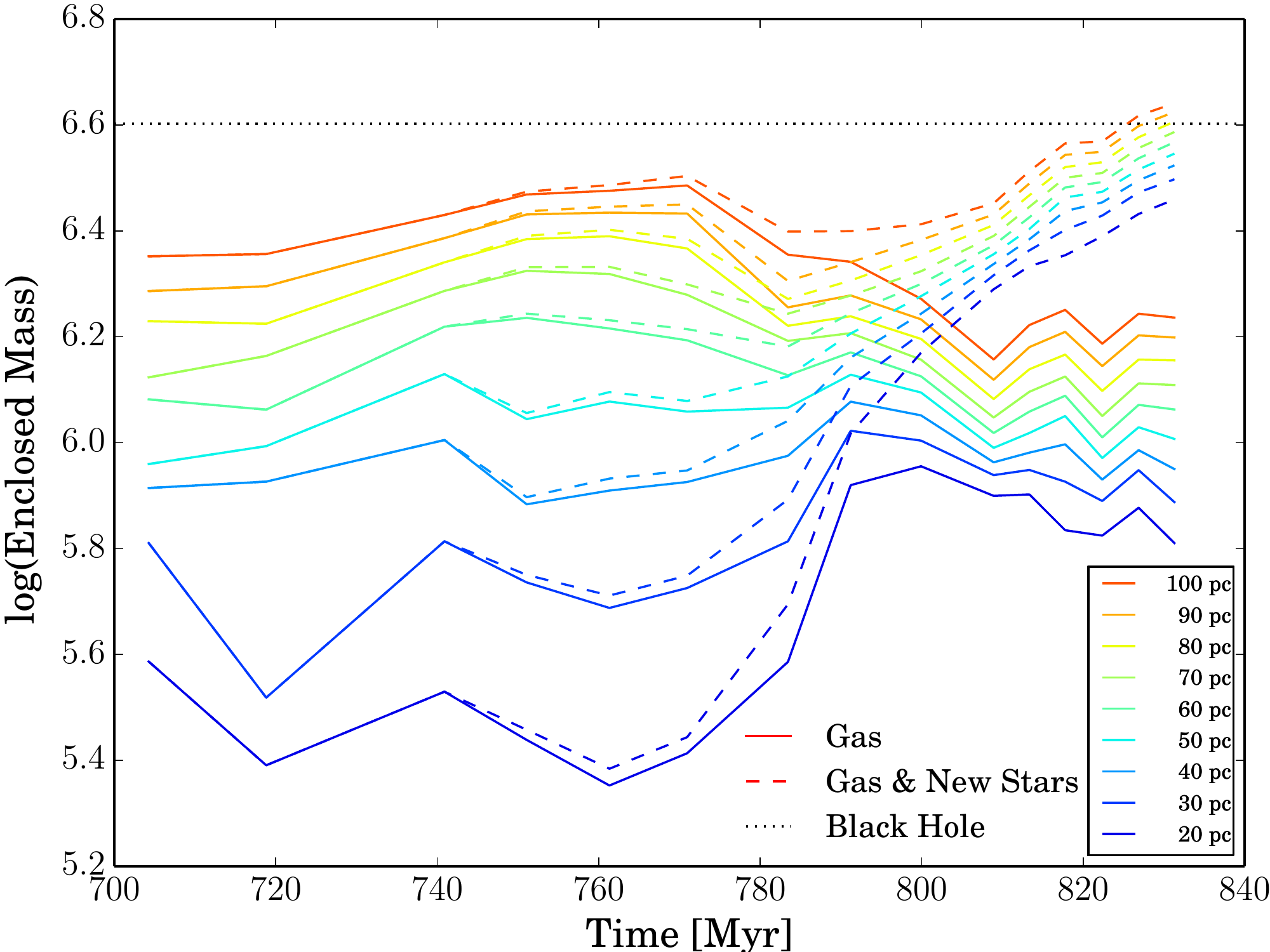}
\caption{Enclosed mass (\Msun\ in log10) within a spherical region centred around the black hole with respect
to time in the simulation: the colour coding corresponds to various radii. The solid lines represent the gas
alone, the dashed lines are for the gas plus stars formed during the simulation, and the dotted black line
indicates the mass of the supermassive black hole, as a reference.}
\label{fig:EnclosedMass}
\end{figure}

The gas distribution evolves rapidly with sometimes three
armlets linking the small and large- scale structure (Fig.~\ref{fig:20Snaps}). The inner Lindblad resonance ring at
40~pc is most visible at $t \sim 750$~Myr, and has an average radial thickness of about
10~pc. Most of the new stars within 100~pc form in that thick ring. At $t \sim 755$~Myr, the wrapping of the
inner spiral forms a secondary elongated ring-like gas distribution with a radius of about 85~pc.
This is quite visible in e.g., the top right panel of Fig.~\ref{fig:3SnapsQuiver} 
where the in-plane velocity vectors are superimposed on the gas density maps.
Within the inner Lindblad resonance ring (40~pc), the gas inflow is almost exactly cancelled by 
the star formation rate of about $10^{-3}$~\Msun.yr$^{-1}$, 
meaning that the gas mass enclosed within that radius stays relatively constant 
until $t = 770$~Myr (Fig.~\ref{fig:EnclosedMass}). Between 50 and 100~pc, the gas inflow
rate is high enough to pursue the building of the central gas mass concentration 
despite new stars being continuously formed. Between $t = 765$ and 770~Myr, the gaseous mass within
100~pc reaches its peak with about $3.1 \, 10^6$~\Msun,  a value similar to the black hole mass itself. 
 

In the present simulation, the ring starts to fragment when about $3\,10^6$~\Msun\ of gas has been gathered
within the central structure (see Fig.~\ref{fig:EnclosedMass}). As most of this gas is confined
within a vertical layer of 10 to 20~pc, it represents nearly 50\% of the background
stellar mass within the same radial extent and height (and is similar to the mass of the black
hole). The break of the regular inflowing gaseous pattern
triggers the formation of about ten clumps most of them embedded in filamentary over-densities. 
The mass of gas and stars within each clump ranges between 0.5 and $1.5\,10^5$~\Msun\ (derived within a radius of 10~pc 
around the density peak of each clump) and an average mass of $10^5$~\Msun. This is a very rapidly
evolving phase, with the clumps individually interacting
with each other and with the background potential (including the black hole). This leads to the 
disruption of most of these clumps on a timescale shorter than 15~Myr, with
$10^6$~\Msun\ of gas being further fueled within the central 20~pc
and almost entirely converted into new stars. By $t \sim 780$~Myr, 
the inner regular disc and ring have disappeared leaving a very concentrated peak of gas and new stars
surrounding the black hole within a radius of about 5~pc.
Such an instability is naturally expected considering that the central gas disc and the new stars
weigh more than the central dark mass itself and become self-gravitating. However, one of the triggering ingredients
of this violent phase is associated with the motion of the black hole, which is emphasised in the next Section.
Such a violent and rapid evolution may naturally lead to a rapidly time-varying fueling of the black hole
itself, hence explain the intermittency in the Active Galactic Nuclei (AGN) activity \citep{Hopkins2010,Gabor2013}.
\begin{figure*}
\centering
\includegraphics[width=\hsize]{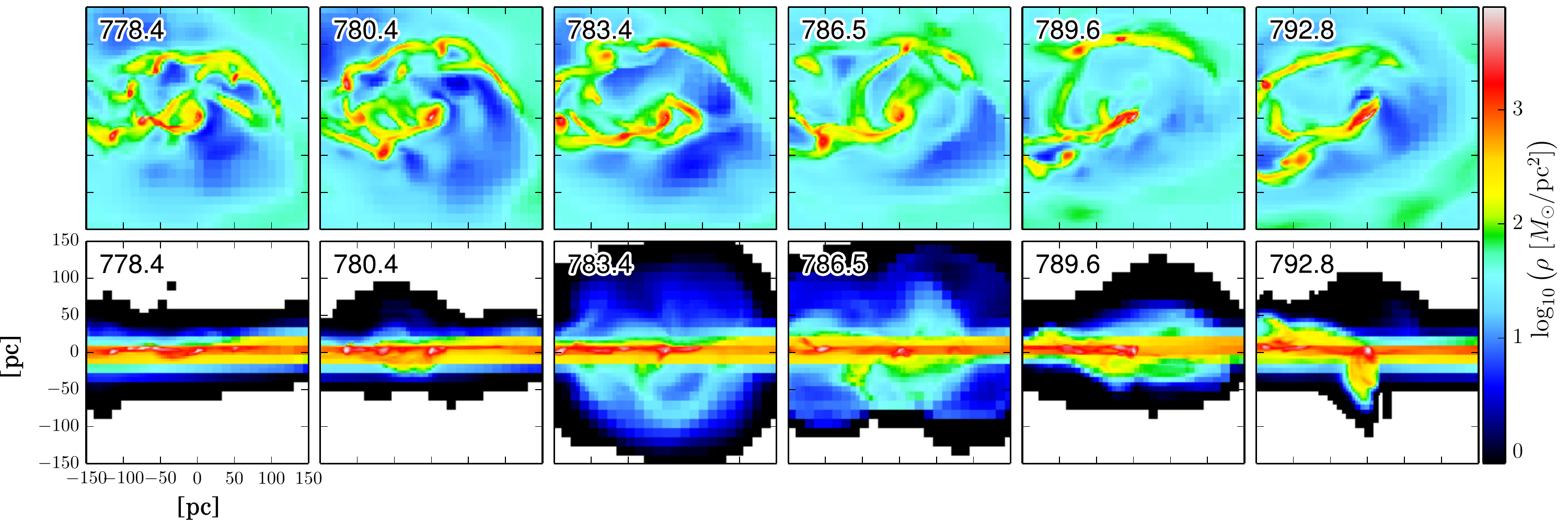}
\caption{Six snapshots over about 15~Myr 
with face-on (first row) and edge-on (second row) views of 
the gas in the central 150~pc. The impact of stellar feedback, which creates bubble-like cavities and filaments 
away from the main plane, is clearly visible.}
\label{fig:Feedback}
\end{figure*}

\subsubsection{The coupling of the black hole and the nuclear cluster}
\label{sec:locking}

As mentioned in Sect.~\ref{sec:overall} and in Fig.~\ref{fig:6gaspanels}, 
the stellar bar is not perfectly symmetric and shows some 
lopsidedness \citep[significant $m=1$ mode, see][and references therein]{Jog2009}. 
The centre of mass of the central kiloparsec does in fact moves around as the bar
rotates, with a maximum amplitude of the order of 50~pc. The central black hole initially follows the centre of gravity
of the inner region of the bar rather accurately until $t \sim 760$~Myr. As gas piles up in the central 
100~pc, star formation starts within a very compact region around the black hole, 
and a stellar cluster slowly forms (at a few $10^{-3}$~\Msun.yr$^{-1}$) 
initially with a radius $R_N$ of less than 10~pc. As a first approximation, we can evaluate the timescale
for the black hole to exchange energy with the cluster by deriving the dynamical friction time
\citep[see e.g.][]{Merritt2013}: 
$T_{df} \sim 5\,10^9 \cdot \left(\sigma_f / 100~{\rm km.s}^{-1} \right)^3 
\cdot \left(\rho / 10^5~{\rm M}_{\odot} {\rm pc}^{-3} \right)^{-1} \cdot \left( m / 10~{\rm M}_{\odot} \right)^{-1} \cdot \left( {\rm ln}
\Lambda / 10 \right)^{-1} {\rm yr}$. For a cluster of $10^5$~\Msun\ within a radius of 10~pc, 
a measured dispersion of 40~\kms, and a black hole of $4\,10^6$~\Msun, $T_{df}$ is shorter than 1~Myr.
The relaxation time $t_R$ of the cluster itself is significantly longer ($\sim 15$~Myr) as it
approximately scales as $t_R \sim \sqrt{R_N^3 / G\,M} \times \left(N / 8 \log{(N)} \right)$, 
where $N$ is the number of particles representing new stars.
The accreted gas mass and the compact stellar cluster newly formed
around the black hole tend to couple these components on a rather short timescale. 
This further tends to decouple the motion of the central gaseous disc, cluster of new stars
and black hole from the centre of gravity of the
older stellar population of the bar which precesses with time.

The decoupling has a direct and significant effect on the gaseous ring as the larger-scale fueling and
resonance structure slowly shifts away from the central black hole, and breaks the symmetry of the orbital
structure and gaseous flow. It can easily be seen in Fig.~\ref{fig:20Snaps}: around $t=767$~Myr,
the ring fragments while the whole gaseous distribution becomes
strongly asymmetric. The increasing offset of the black hole further amplifies the violent evolution of the
gas structure described in the previous Section. It is thus interesting to note that after $T\sim765$~Myr 
the formation of clumps within the central 100~pc is in fact initially the
consequence of the black hole and cluster offsetting, and not solely from the local gas over-density and 
self-gravity.

\subsubsection{Star formation in clumps, and stellar feedback}
\label{sec:clumps}

Star formation is triggered in the densest regions, which include a few of these clumps, but also 
the filamentary overdensities which represent the left-overs of the inner Lindblad ring.
The star formation in these clumps sums up to about $2\,10^{-3}$~\Msun.yr$^{-1}$ within 100~pc.
The clumps are organised within a twisted elliptical structure
with a diameter of about 250~pc, and a vertical extension of 20~pc, 
the black hole being offset by about 70~pc from its centre (Figs~\ref{fig:20Snaps} and ~\ref{fig:Feedback}).
As mentioned, the following fast evolution tends to both disrupt these structures, and make them merge
within the central 20~pc. By $t\sim780$~Myr, only a few clumps remain as all have 
been disrupted or merged with the very central overdensity.
This triggers a rapid density increase after $t=790$~Myr and a subsequent ten fold increase
of the star formation rate: it is clearly visible in Fig.~\ref{fig:EnclosedMass} as the 
mass of gas enclosed in 20 and 100~pc respectively first decreases and then increases rapidly: 
in the later step the mass within 20~pc goes from $\sim8$\% to $\sim 45$\% of the mass within 100~pc
(excluding the black hole).
The inflow is associated with a net increase of the star formation rate which reaches 0.05~\Msun.yr$^{-1}$,
almost entirely focused in the very close environment of the black hole.
This is to be compared with the star formation rate at the end of the bar of a few \Msun.yr$^{-1}$.
The violent evolution of the gas distribution also constrains the availability of gas directly fueling
the black hole: this could be a natural cause of AGN variability in galaxies on timescales of a few Myr.

\begin{figure}
\centering
\includegraphics[width=\columnwidth]{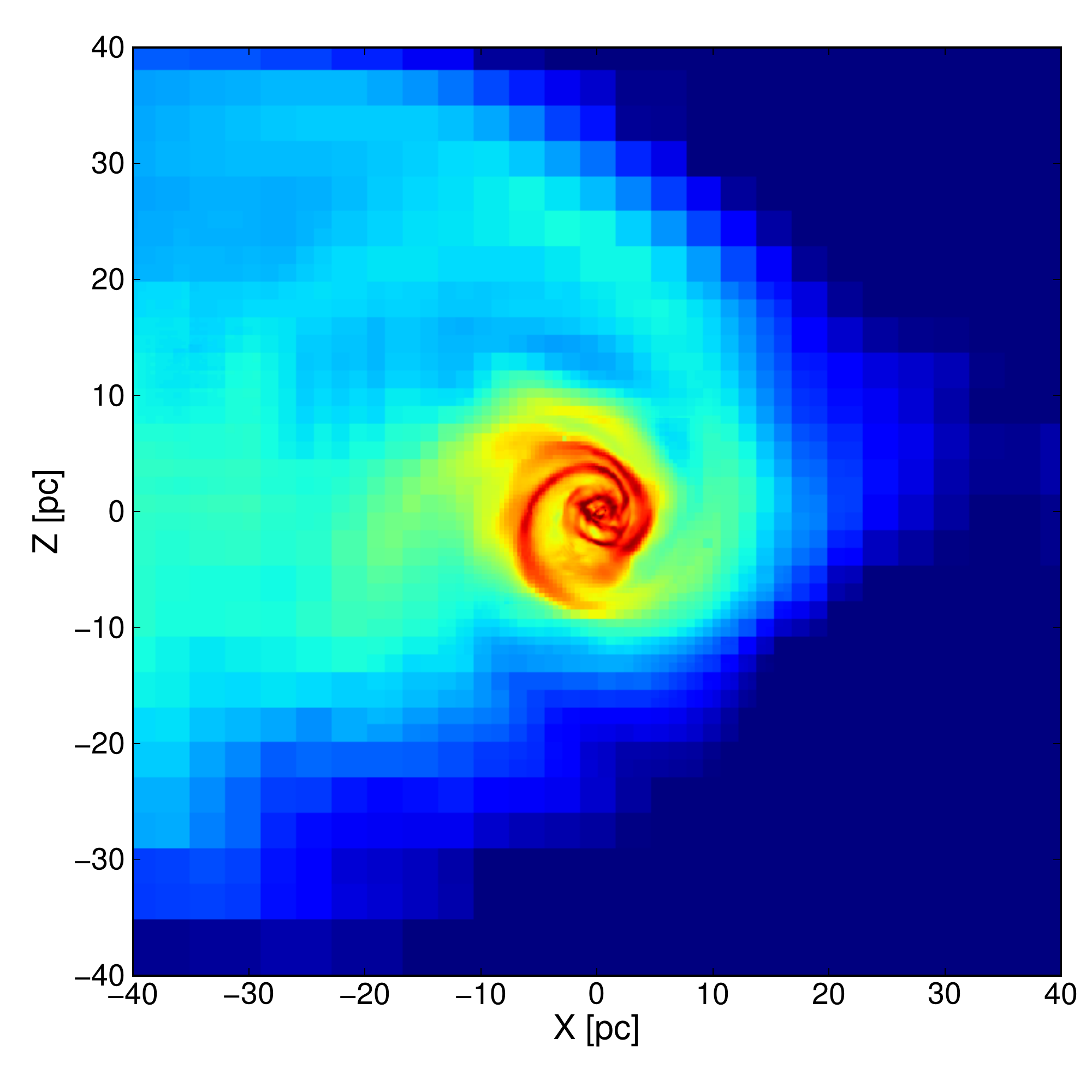}
\caption{Nearly face-on view ($X, Z$) of the accretion within the 40~pc around the black hole at $t=830$~Myr:
note that this view corresponds to an edge-on view of the main galaxy plane ($X,Y$).}
\label{fig:faceOnDisk}
\end{figure}

Another important ingredient for the evolution of the central structure is the feedback associated with the
starburst in the clumps. Following the prescriptions implemented in the simulation, this feedback concerns
20\% of the formed stellar mass. It first takes the form of a photo-ionisation bubble and radiative pressure,
followed by supernova explosions (via injection of kinetic energy) activated 10~Myr after new stars are formed. The stellar feedback expels gas outside the main plane and up to about 200~pc as illustrated in Fig.~\ref{fig:Feedback}. It creates a large expanding bubble (see third snapshot in Fig.~\ref{fig:Feedback}), the gas ultimately falling back onto the disc in a few Myrs, part of it being caught in the environment of the black hole. A rather thin nearly vertical gaseous 
disc with an extent of about 20~pc is thus being formed (Fig.~\ref{fig:faceOnDisk}), which continues to slowly
accrete gas and form stars. Gas clouds being accreted close to the black hole are tidally stretched during their course\footnote{See
the extreme case of a highly stretched gas cloud around the Milky Way black hole in \cite{Gillessen2013}.},
creating a series of extended tidal-like features (from 5 to 50~pc in size) 
precessing within the vertical disc: this can be
seen in Fig.~\ref{fig:faceOnDisk} but is best illustrated when only the dense gas is shown as in Fig.~\ref{fig:Sketch}.
These filaments may be associated with the streamers observed at the centre of the Milky Way (see
discussion in Sect.~\ref{sec:MWbar}).
The stellar-driven feedback thus helps remove angular momentum from the gas, and allows it to come closer
to the supermassive black hole.
In this regard, it is important to notice that it is the asymmetry of gas infall after
star formation feedback, which produces the lopsidedness of the gas distribution
in the center of the galaxy: this was already alluded to by \cite{Rodriguez-Fernandez2008},
who argued that the asymmetry of the central potential is insufficient to account for the observed lopsidedness.

\begin{figure}
\centering
\includegraphics[width=\columnwidth]{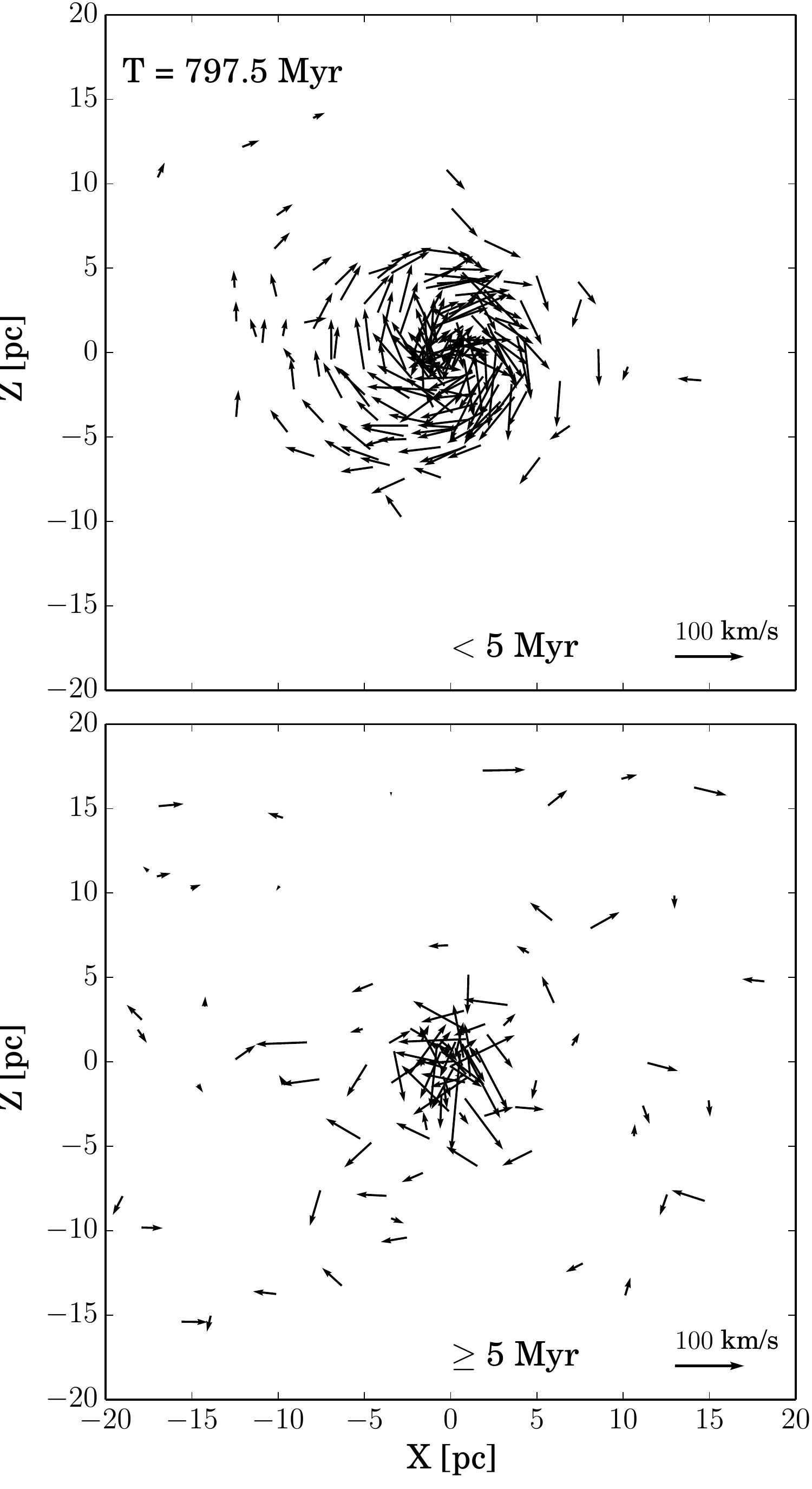}
\caption{Inner $X,Z$ view of the (formed) stars positions and velocities in 
the central 20~pc around the black hole at $t\sim800$~Myr. Arrows correspond
to the velocities in the $X,Z$ plane (the velocity scale is indicated at the bottom right). The top 
(resp. bottom) panel corresponds to more recently, $< 5$~Myr, (resp. earlier, $> 5$~Myr) formed particles.}
\label{fig:Quiver}
\end{figure}

\subsubsection{The nuclear star cluster}

\begin{figure}
\centering
\includegraphics[width=\columnwidth]{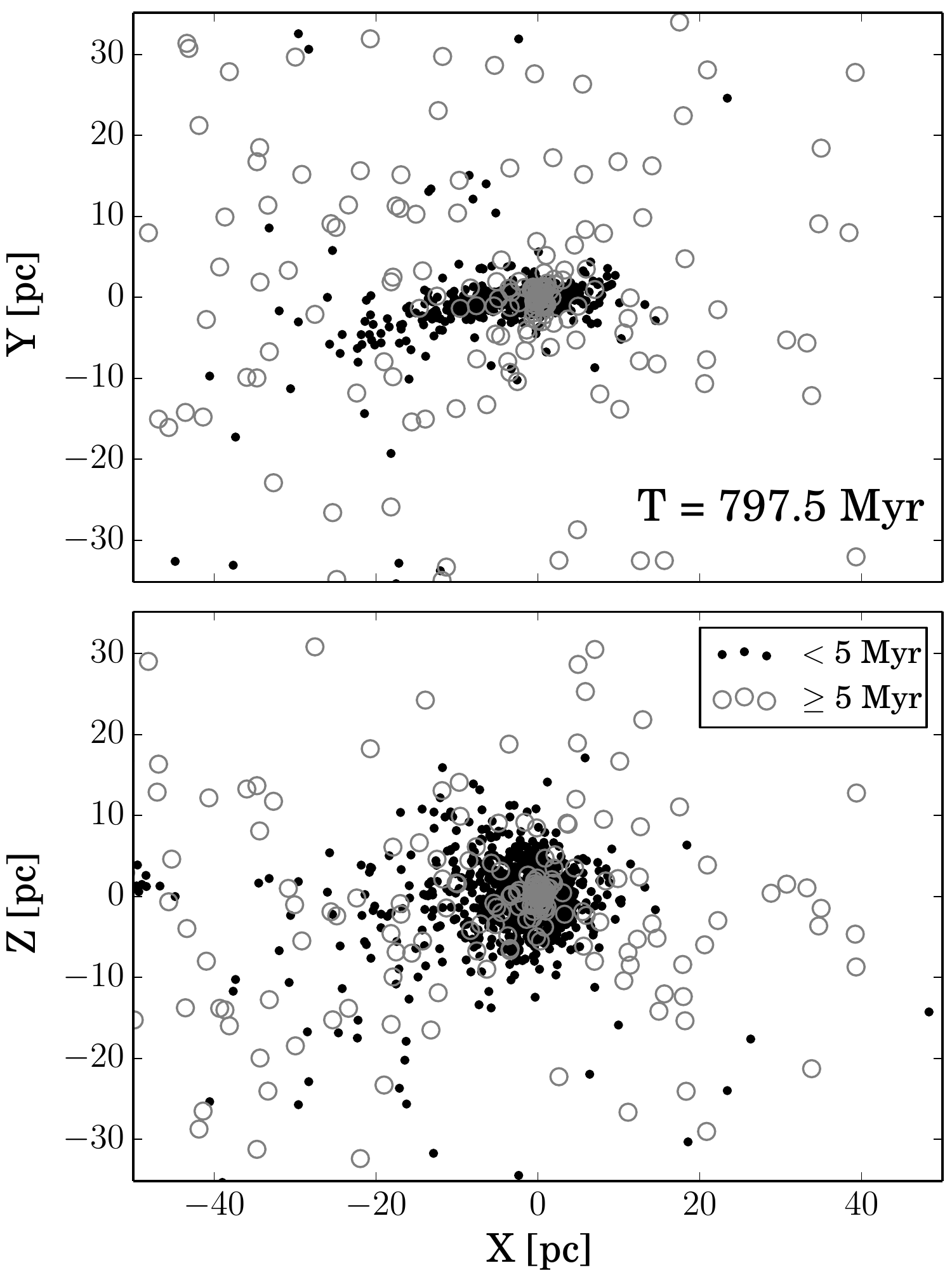}
\caption{$X,Z$ and $X,Y$ views of the stars formed during the simulation
within 40~pc at $t\sim800$~Myr, splitted in two age bins ($< 5$~Myr, black dots; $> 5$~Myr, open grey circles). 
The very young stars are mostly
distributed within a disc-like structure which is perpendicular to the main galaxy plane ($X,Y$). Note the
core of stars older than 5~Myr, and its more extended and flattened distribution.}
\label{fig:DistribNS}
\end{figure}

\begin{figure}
\centering
\includegraphics[width=\columnwidth]{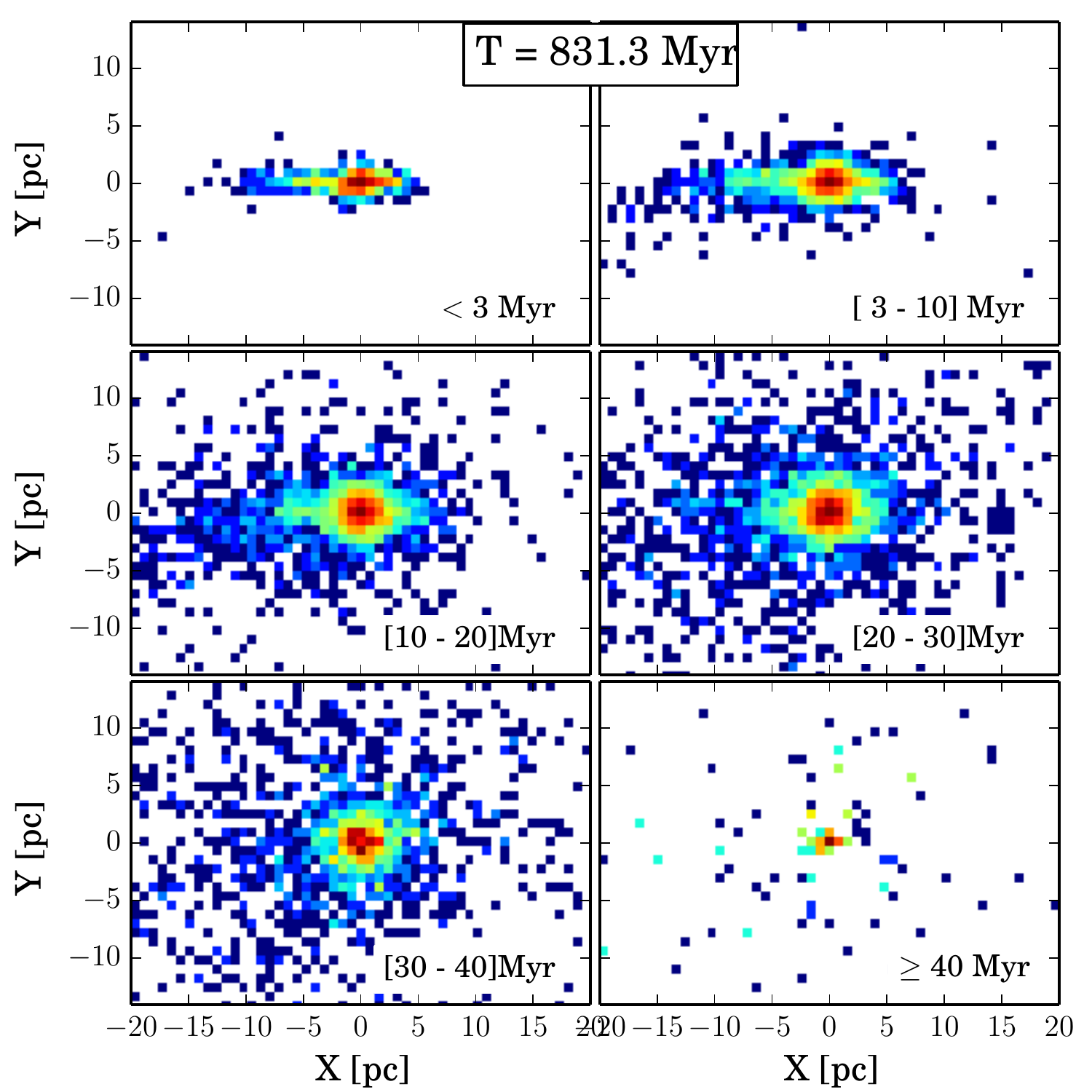}
\caption{$X,Y$ views of the stars formed during the simulation within 20~pc at $t\sim830$~Myr separated by ages. Panels
go from younger than 3~Myr (top left) to older than 40~Myr (bottom right).}
\label{fig:AgeNewStars}
\end{figure}

The above-mentioned evolution involves several subsequent episodes of star formation around the black hole. 
During the first phase ($t \lesssim 760$~Myr), 
stars forming within 100~pc of the black hole are the result of the radial
inward motion of the gas associated with the resonance ring, building a nuclear star cluster
of a few $10^5$~\Msun. Most of these stars have thus an overall angular momentum vector aligned with 
the one of the main galaxy disc. Only stars within the inner 5~pc have a more isotropic
motion, following the nearly spherical potential imposed by the black hole.
In a second phase, after $t\sim790$~Myr, new stars form from gas which has been expelled via stellar-driven
feedback and been re-accreted in a gaseous disc-like structure, perpendicular to the galaxy plane. 
These new stars have a well-organised rotation with velocities going from 30 to 130~\kms\ 
between 3 and 10~pc (see Fig.~\ref{fig:Quiver}), and velocity dispersions between 8 and 15~\kms.
Within the central 2~pc, we observe more counter-rotating stars
and the dispersion becomes larger than the mean azimuthal velocity.
The gas distribution and velocities at a scale of 10~pc exhibit an off-centring (see Figs.~\ref{fig:faceOnDisk} 
and \ref{fig:Quiver}) which can be associated with an
$m=1$ mode, naturally expected for a potential dominated by a point mass \citep[e.g.][]{Masset1997, Hopkins2010}.
The averaged angular momentum vector of this inclined disc lies within the large-scale galactic plane. 
The newly formed stars are thus superposed with the slightly older, hotter and more isotropic population of stars
formed in the first phase as illustrated in Fig.~\ref{fig:DistribNS}.
At the end of the simulation ($t \sim 830$~Myr), the nuclear star cluster is therefore made up 
of several generations of stars, the youngest having an average angular momentum vector oriented
within the galaxy plane (perpendicular to the total angular momentum vector of the galaxy), 
while the older have a more spherical and isotropic distribution (Figs.~\ref{fig:DistribNS} and ~\ref{fig:AgeNewStars}).
This is reminiscent of the recent observations by \cite{Feldmeier2014} who witnessed the potential
presence of a set of stars orbiting perpendicularly with respect to the Galactic plane.
More generally, it may lead to a natural explanation for multiple stellar populations with
different kinematic signatures in the central regions of galaxies, even though
the time window within which such signatures can be observed should be addressed.


\section{Discussion and Conclusions}
\label{sec:conclusions}
\begin{figure*}
\centering
\includegraphics[width=15cm]{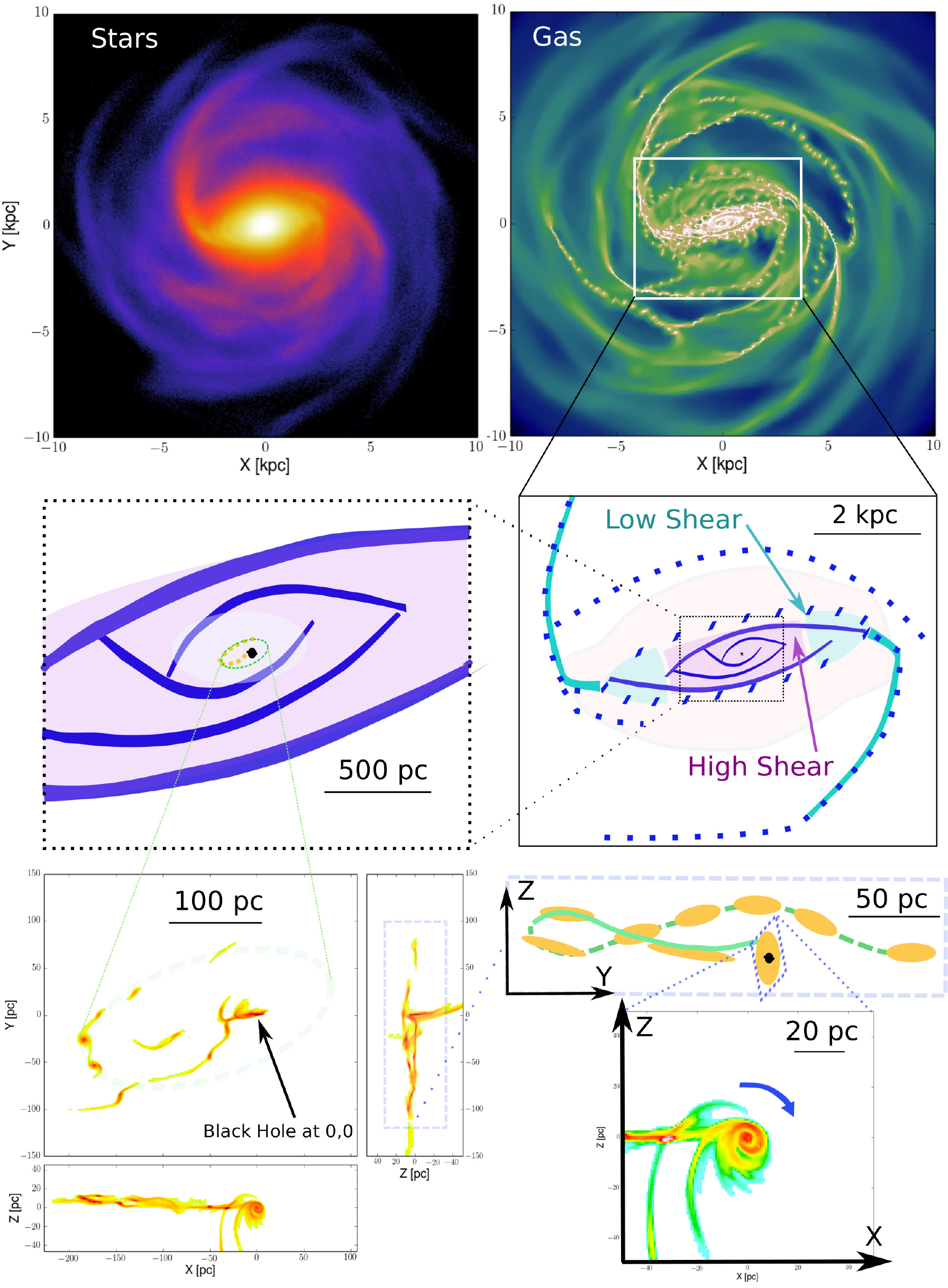}
\caption{Combined set of sketchs and maps emphasising the structures observed in the simulation at various scales.
Top panels: the stellar (left) and gaseous (right) mass distributions within 10~kpc, seen face-on.
Middle panels: sketches illustrating the structures, including star forming clumps (blue dots), gas spirals
and lanes (green and purple curved lines). The background colors delineate the bar (pale red), and regions
with low and high shear within the bar (as in Figure~\ref{fig:Shear}).
Bottom left panels: face-on ($X,Y$), edge-on ($X,Z$) and end-on ($Y,Z$) 
views of the gas within the central $\sim200$~pc emphasising the dense gas clouds and their asymmetric
distribution at $t\sim800$~Myr. Note also the vertical extent of the ring-like structure seen in the edge-on
and end-on view. Bottom right panels: illustration of the torus-like ring and its clumps of dense gas
connected to the vertical disc surrounding the black hole.}
\label{fig:Sketch}
\end{figure*}

\subsection{Chronology and structures}

We now summarise the chronology of events observed within the simulations and mentioned in the last Sections, 
to emphasise a few important ingredients and processes. It is important to emphasise agian here that
most detected features within the bar are rapidly evolving and thus transient.
\begin{itemize}
    \item Despite the presence of a significant amount of gas within the bar region, star formation 
        proceeds mostly at its ends, or within a radius of about 100~pc near the black hole and not 
        in between;
    \item Resonant structures (Lindblad) focus gas in elongated ring-like distributions, themselves connected
        to the central dark mass via mini-spirals within a few tens of parsecs;
    \item Gas clumps of a few $10^5$~\Msun\ are rapidly forming as a consequence of the gas accumulation and
        symmetry breaking described in Sect.~\ref{sec:locking}. 
        However, only the most massive clumps manage to efficiently form stars, and they rapidly get disrupted;
    \item Star formation and stellar feedback are important ingredients which allow 
        the gas to be transported at higher heights, part of it falling back to be 
        accreted into a dense structure around the black hole. AGN feedback could in principle play
        a comparable role.
    \item The resulting inner elliptical (ring-like) structure with a diameter of about 250~pc 
          exhibits a vertical extension of 10 to 20~pc, the black hole being offset by about 70~pc from its
          centre.
    \item Gas accreted around the black hole within a radius of about 10~pc is distributed in a disc-like
        structure decoupled from the main Galactic plane, with its angular momentum vector 
        aligned within the main galactic plane.
    \item The gas accreted in the vicinity of the black hole is stretched by the associated tidal forces, creating
         a series of long winding tails of gaseous material.
\end{itemize}

To better illustrate the corresponding scales and structures, we use $t \sim 795$~Myr as a reference
time and present the actual stellar and gas distributions with sketches flagging
the most important features in Fig.~\ref{fig:Sketch}. At a scale of a few kiloparsecs, the bar and associated
spiral arms dominate. Star formation mostly proceeds in the spiral arms \citep[see][for details]{Renaud2013,
Kraljic2014} and at the ends of the bar, as discussed in Sect.~\ref{sec:lack}.
A series of intertwined gas filaments connect the ends of the bar down to the central 250~pc where the gas
partly ends in a thick pseudo-ring surrounding a lower density region. The gas structures continue inwards
with thinner spirals arms connecting with the central 50~pc ring and disc seen in Fig.~\ref{fig:20Snaps}.
In the simulation we present here, about 2~10$^7$~\Msun\ of gas is
accumulated within a radius of 200~pc around the black hole, a few 10$^6$~\Msun\ within 100~pc (see
Fig.~\ref{fig:EnclosedMass}), and about 2~10$^5$~\Msun\ within the 10~pc radius vertical disc. 
As emphasised in Fig.~\ref{fig:EnclosedMass}, the gas mass within the 100~pc stays relatively constant as
star formation proceeds. 
The accelerated fueling rate in the central 200~pc occurs when the instability sets in, thus breaking the
ILR barrier \citep[see also,][]{Englmaier2004}. Hence it is the combination of the bar-driven fueling, the lack of star formation in the bar,
and the onset of the instability at the centre, which manages to bring gas in the close vicinity of the black hole: these 
three processes are thus followed simultaneously within the present simulation which then leads to the
formation of a polar disc at a scale of 10~pc around the black hole.

\subsection{Limitations of the simulation}

It is important to probe whether the results and processes described above are 
specific to the considered simulation, with its characteristics (e.g., spatial and mass resolutions, numerical 
scheme) and implemented recipes (star formation, stellar-driven feedback). 
We thus briefly review processes which may be significantly affected by varying prescriptions.

Even though the morphology and evolution of resonance rings
certainly depend on the details pertaining to the mass profile, gas content, etc,
the large-scale fueling of gas towards resonances is a generic property of 
bars \citep{Athanassoula1992,Combes2001}. In the present
simulation, the existence of a central spiral fueling the inner region is associated with the
presence of a central dark mass as theoretically expected \citep[see e.g.][]{Fukuda1998}
and observed \citep{Combes2013, Combes2014, Garcia-Burillo2014}. 
The exact evolutionary steps followed by the gas
distribution at the very centre (Fig.~\ref{fig:20Snaps}) are nevertheless closely linked to the star formation and
feedback implemented schemes. It may also heavily depends on the prior presence of a massive stellar cluster
around the black hole, as such a seed cluster would imply a more diffuse mass distribution and a consequently
less peaked velocity gradient, at least in the central few tens of parsecs.

The relative impact of shear on the local gas densities inside the bar (Sect.~\ref{sec:lack},
Fig.~\ref{fig:Shear}), at its ends
and outside may vary, both in terms of the properties of the bar itself and the physical ingredients 
included in the simulation and its detailed implementation.
In the inner regions, it seems, however, hard to prevent high shear for medium
to high strength bars considering that it is a direct consequence of the orbital 
structure in the bar combined with the dissipative nature of the gas. 
Measurements of shear for clouds within actual bars would help confirm this picture. 

Another set of limitations originates in the treatment of the non-dissipative particles, namely the old
stars, dark matter and the massive black hole. 
These are evolved on a sub-grid with an associated spatial softening of 3~pc, hence coarser than the
grid following the very dense gas regions. New stars form with a mass of a few hundreds solar masses, thus
not sampling a given initial mass function but always representing an indissociable group of simultaneously
born stars. Additionally, 
the cartesian (adaptive) grid itself may impose some alignment in the observed structures which
may partly impact on the vertical orientation of the central disc structure around the black hole
\citep{Hahn2010,Dubois2014}. However, the vertical structure is mostly 
set by the angular momentum of the incoming gas:
the nuclear disc is in fact not initially aligned with the grid main axes, and its symmetry axis shifts with
time, going away from the grid orientation. 
At the end of the simulation, most of the mass in that disc is concentrated within a radius of 10~pc,
about 200 times the smallest resolution element and gas is being accreted from larger scale 
filaments which are not aligned with the inner disc.
Last but not least, supernovae feedback is provided at a constant rate which depends directly on the mass of new
stars, and is not described as a stochastic event. 
These ingredients certainly impact on the interaction between e.g., 
the dense gas clumps and the new stellar populations. It thus affects the relaxation at the very centre 
between the new stars and the black hole. This may consequently induce a different evolutionary path for the
central region: the importance of each event in the chronology described above should therefore be probed via
different simulations starting with various prescriptions and initial conditions.

\subsection{Other barred galaxies and our Milky Way}
\label{sec:MWbar}

We finally wish to relate the various features and evolution 
witnessed in the simulation with structures actually observed in our own Milky Way or in external galaxies. 
We first start by briefly pointing out the lack
of star formation in the inner part of the bar (Sect.\ref{sec:lack}).

The classification of barred galaxies following the spatial distribution of their 
H${\tt II}$ regions generally leads to two or three groups
\citep{Phillips1996,Verley2007}. Various authors thus emphasised the lack 
of star formation within the bar region for early-type galaxies (SBb ad SBa) with strong bars.
An early study of NGC\,1530 beautifully shows such a concentration of star forming regions solely at
the ends of the galactic bar or around the nucleus \citep{Reynaud1998}: the authors argued for the role
of bar-induced shear as the source for the disruption of potential star forming clouds.
The lack of star formation within the bar region despite the presence of diffuse gas
therefore seems to be a rather common property of disc galaxies with strong bars.
\cite{Phillips1996} and \cite{Martin1997}
also point out that these systems often exhibit a starbursting 
nuclear region very probably associated to inner resonances,
a prototypical example being NGC\,1300 \citep[see also][and references therein]{Elmegreen2009}.  

In our own Galaxy, studies of actual Galactic giant molecular clouds like W43 \citep{Carlhoff2013, Motte2014} 
seem to indicate
low shear values for dense clouds lying close to the end of the Galactic bar, where heightened star formation
occurs. On the other hand, \cite{Longmore2013a} pointed out the very low level of star formation
down to a radius of about 500~pc. The inner 500~pc region, labelled Central Molecular Zone (CMZ), seems to
harbour high densities of gas, but with very little star forming sites. \cite{Longmore2013a} and
\cite{Kruijssen2014} have
thus argued that the star formation rate within the CMZ is significantly below what is expected considering
the amount of high-surface density gas. 

Going further inward, the inner dense molecular structure is often described as an elliptical and twisted
ring of semi major-axis 100~pc \citep{Molinari2011}. SgrA seems to be offset from this ring
by about 50~pc. The distribution of dense gas is not symmetric along that ring as emphasised by recent
maps of cold dust and gas clouds \citep[e.g.][]{Molinari2011,Liu2013}. The presence of gas streamers
was also recently reexamined by \citep{Liu2012}, who nicely illustrated (see their Fig.~7) the connection
between a central molecular ring a few parsecs in size and larger-scale gas filaments which seems to be
tidally distorted and coiled. 

All these features are qualitatively reminiscent of the ones observed at the 200~pc and 10~pc scales in the
present simulation (see Figs.~\ref{fig:faceOnDisk} and \ref{fig:Sketch}): 
the twist of the ring formed by the few clumps at a scale of about 150~pc, distributed
asymmetrically around the black hole, which itself is surrounded by accreted infalling gas, some of it being
stretched by tidal forces. The fact that some of the scales are similar (the size of 100~pc ring, the offset
of the black hole) is certainly a coincidence. However, the fact that such structures naturally emerge
from relatively generic processes (gas accretion, dynamical friction and dissipation, star-driven feedback)
hints for new avenues to probe the formation and evolution processes of galactic nuclei.
We should also emphasise the case of M\,83 \citep{Thatte2000, Diaz2006, Houghton2008} which hosts a bar resembling the one in
the present simulation, and has an intriguing asymmetric circumnuclear region. The latter has been
interpreted as the signature of a transient $m=1$ mode \citep{Knapen2010}, the result of 
a complex startburt episode \citep{PiquerasLopez2012}, or of an accretion event \citep{Rodrigues2009}: the
observed central structures in M\,83 could thus be reconsidered in the light of the simulation we presented in this paper.

At the end of the simulation described in the present paper, the bar is still slowly
growing. It is therefore impossible to robustly address its long term evolution. 
Nevertheless, one of the most important characteristics of the features observed in the simulation described above
is their transient nature: redistribution of gas and its fuelling towards the central region, building up
of the mass in the vicinity of the black hole, star formation and stellar-driven feedback all
occur within a time window shorter than 100~Myr. 
The central structures significantly evolve on an even shorter timescale, 
namely a few Myr, under the strong influence of instabilities and 
the motion of the black hole. This could be a natural source of intermittency for the fueling of the black
hole itself, hence partly causing the time-varying nature of AGN \cite{Hopkins2010,Gabor2013}. 
We could also speculate that after the violent phase which
resulted in the formation of a nuclear (vertical) disc around the black hole, the gas flow driven by the bar
tumbling potential would resume. This could be the starting point for a second cycle with in-spiraling gas
being accumulated in a ring or disc-like structure at a scale of 100-200~pc, hence triggering another
episode of rapid evolution where the black hole, and stellar-driven feedback would play a major role. 
Such a two-phase scenario is clearly reminiscent of the one proposed in a pioneering paper by \cite{Shlosman1989},
where gas is first transported from kpc down to pc scales via dynamical instabilities, accumulating mass within an inner
gaseous disc which could then become unstable and provide fuel to an existing black hole (or form one).
Qualitatively similar cycles have also been proposed by e.g., \cite{Kruijssen2013} as a potential mechanism to explain the
rather low efficiency of star formation in the CMZ. We emphasise here the critical roles of the central
dark mass and of the symmetry breaking for the evolution of the structures,
as well as the importance of feedback for the redistribution of angular momentum. 

The present study thus illustrates the importance of the interplay between large-scale 
and small scale dynamics, and the potential richness
initiated by the presence of a bar, a black hole, gas dynamics, star formation and the associated feedback.
If a second cycle were to be triggered, a number of key ingredients would naturally differ
in the initial conditions. The potential of the bar itself is changing with
time, and is being influenced by the exchange of angular momentum between the stars and the gas
\citep{Bournaud2005}. The central mass distribution and dynamics should also be significantly different, and
the central black hole may be expected to grow in mass (something we did not follow in the present simulation). 
Still, the formation of resonant structures at scales
of about 100~pc and smaller, followed by gravitational instabilities, the formation of gas clumps, 
star formation, outflow and re-accretion closer to the black hole could be generic features for
gas-rich spiral galaxies. 
If such cycles occur, as already nicely emphasised by e.g., \cite{Stark2004}, 
our simulation suggests that they would span time periods of $\sim 100-200$~Myr.
Their detailed properties and evolutionary tracks will obviously depend on many ingredients 
including the relative mass of the black hole and the availability of gas.
This would certainly condition the mode of accretion onto the black hole, but also
its time variability.

Similar time-varying structures have been witnessed in simulations by \cite{Hopkins2010}
\cite{Levine2008}, or \cite{Mayer2010}. \cite{Hopkins2010} used ``zooming-in'' techniques
to reach sub-parsec scales in a series of simulations varying the gas fraction
and treatments of the gas feedback. Some of their simulations develop one-arm spirals with a clumpy
structure \citep[namely Nf8h1c0 in][]{Hopkins2010a}, but these are for very gas-rich systems (mass fraction of
gas within 100~pc of about 0.5) with high accretion rates of a few \Msun.yr$^{-1}$ inside 100~pc, so that the
clumpiness directly derives from the gas self-gravity. Overall, the simulations presented and discussed in \cite{Hopkins2010a, Hopkins2010} 
show a rather smooth gaseous distribution within the central 100~pc, but as emphasised by \cite{Hopkins2010a}, this heavily
depends on the implemented stellar-driven feedback recipe, as well as on the imposed cooling floor.
\cite{Levine2008} also used a zoom-in technique to reach cell sizes as small as 0.03~pc in a  simulated
disc galaxy within a cosmological volume. A range of
instabilities and spiral structures are observed from kpc to pc scales. 
A circumnuclear self-gravitating gas disc forms during the simulation and has a remarkably constant 
power-law density profile. Both \cite{Hopkins2010} and \cite{Levine2008} depict a situation where
large-scale gas inflows are orders of magnitude higher than in the simulation we describe in the present paper. As for the
study of \cite{Mayer2010}, a merger-driven gas inflow is followed with a spatial resolution down to a tenth of a parsec:
the onset of a unstable massive disc around the supermassive black hole is even more rapid, 
hence a very high corresponding accretion rate. 

The main specific attribute of the simulation we present here
is then the relatively quiet accretion mode at the 100~pc scale ($\leq 0.02$~\Msun.yr$^{-1}$), even though we witness a rapidly
evolving structure within the central 200~pc. We self-consistently follow this evolution (while ignoring
any potential AGN feedback process) with the formation of a decoupled 10~pc scale (polar) circumnuclear structure which may
be of high relevance for the evolution of galactic nuclear regions.
In that context, high resolution hydrodynamical simulations specifically 
focusing on few parsecs domains have also been used to study the interaction between 
infalling gas clouds and a black hole \citep[see e.g.][]{Lucas2013,Alig2013} in an attempt to 
better understand the complex stellar and gaseous morphologies and dynamics observed near the Galactic centre.
More recent constraints on their formation processes come from the observations of e.g., \cite{Feldmeier2014}
who managed to map the detailed stellar kinematics of the central cluster, and suggest the existence of a
stellar perpendicular structure within the inner 1~pc.
Considering the connections and coupling between the various scales and processes, as emphasised in the
present paper, it would thus be important to examine such issues with self-consistent
simulations at very high mass and spatial resolutions.

\section*{Acknowledgements}
We would like to warmly thank Yohan Dubois and Romain Teyssier for useful input which helped the writing of 
this paper. We thank Diederick Kruijssen for his input on the final stages of the manuscript. 
F. Renaud, F. Bournaud and J. Gabor acknowledge support from the EC through grant ERC-StG-257720. 
This work was granted access to the PRACE Research Infrastructure resource Curie hosted at CEA-TGCC, 
France (PRACE project ra-0283-, and national GENCI resources, projects 2013-GEN2192 and 2014-GEN2192). 
\bibliography{MilkyWay}{}
\bibliographystyle{mn2e_fixed}

\end{document}